\documentclass[11pt]{article}	
\usepackage{amsmath,amsthm,amsfonts,amscd, amsmath}  
\usepackage{amssymb,amsfonts,amsmath,amscd,amsthm}	
\usepackage{graphicx}
\usepackage{setspace}
\usepackage[caption=false]{subfig}    
\usepackage{verbatim}      	
\usepackage{makeidx}       	
\usepackage{color}
\usepackage{enumerate}
\usepackage{authblk}
\usepackage{epsfig}         	
\usepackage{url}		
\usepackage{bm}

\usepackage[super,comma,sort&compress]{} 

\usepackage{hypernat}



\theoremstyle{definition}

\theoremstyle{remark}

\newcommand{\latexe}{{\LaTeX\kern.125em2%
                      \lower.5ex\hbox{$\varepsilon$}}}

\chardef\bslash=`\\	

\makeatletter		
\def\square{\RIfM@\bgroup\else$\bgroup\aftergroup$\fi
  \vcenter{\hrule\hbox{\vrule\@height.6em\kern.6em\vrule}%
                                              \hrule}\egroup}

\makeatother		
\makeindex    

\usepackage{fancyhdr}
\newcommand{\blue}[1]{\textcolor{black}{#1}}

\begin{document}
\title{A Probabilistic Design Method for
Fatigue Life of Metallic Component
}

\author[,1]{Danial Faghihi\thanks{Corresponding author. Research Associate, Institute for Computational Engineering \& Sciences,
The University of Texas at Austin, Austin, Texas 78712.
Email: danial@ices.utexas.edu. Tel.: +1 512 232 7219. }}
\author[2]{Subhasis Sarkar}
\author[2]{Mehdi Naderi}
\author[3]{Lloyd Hackel}
\author[2]{Nagaraja Iyyer}

\affil[1]{\small{Institute for Computational Engineering and Sciences,

The University of Texas at Austin}}
\affil[2]{Technical Data Analysis, Inc., Falls Church, VA}
\affil[3]{Curtiss-Wright, MIC-Laser Peening Division, Livermore CA}

\date{\today}
\maketitle


\begin{abstract}

In the present study, a general probabilistic design framework is developed for cyclic fatigue life prediction of metallic hardware using methods that address uncertainty in experimental data and computational model. The methodology involves
(i)	fatigue test data conducted on coupons of Ti6Al4V material
(ii)	continuum damage mechanics based material constitutive models to simulate cyclic fatigue behavior of material
(iii)	variance-based global sensitivity analysis
(iv)	Bayesian framework for model calibration and uncertainty quantification and
(v)	computational life prediction and probabilistic design decision making under uncertainty.
The outcomes of computational analyses using the experimental data prove the
feasibility of the probabilistic design methods for model calibration in presence of incomplete and noisy data.
Moreover, using  probabilistic design methods result in assessment of reliability of fatigue life predicted by computational models.

\end{abstract}

\noindent \textit{Keywords}:
Probabilistic design,
cyclic fatigue life prediction,
continuum damage mechanics,
Bayesian model calibration,
sensitivity analysis.



\section{Introduction}\label{sec:intro}
\blue{
Recent advances in computational simulations of materials and mechanical systems have increased employment of simulation-based engineering design that relies on computational models to predict component performances. Such approach allows for decision-making prior to the availability of physical prototypes and provides insights into certain phenomena for which experimental observations are not available due to measurement limitations. Despite these benefits, the expected question that arises is the reliability of such decision-making as a means of predicting physical reality. This is an important question, as there is uncertainty in every phase of the design process of engineering systems \cite{OdenMoserGhattas2010I,OdenMoserGhattas2010II} including manufacturing imprecision, variation of material properties, and incomplete or lack of experimental measurements. Such uncertainties might have significant impacts on design performance and dealing with them is one of the emerging challenges in product development and is the main bottle neck in simulation-based engineering design.
Conventional (deterministic) design methods are inadequate to cope with uncertainties and results in invalid decision-making and design with unknown reliability or unknown risk. Probabilistic engineering design, on the other hand, relies on probabilistic and statistical methods to assess reliability within a design criterion. This allows for probabilistic risk assessment to be built into the design process and provides means for quantification of uncertainties and their mitigation in the design stage. In this regard, probabilistic design methods are essential for modern engineering that enables solving complex engineering problems in the face of uncertainty.
}

\blue{
The prediction of life of mechanical components such as turbine engines experiencing cyclic loading is of vital importance to aircraft propulsion systems and their mission capability. Fatigue cracking and failure of turbine engine components are typically initiated in the areas of high mechanical stress. Surface treatments, such as laser peening, enables engineering of protective pre-stress into these high stress areas, and have been proven to be of great benefit to extend component lifetimes and mitigate the safety-related risk of high cycle fatigue.
Continuum damage mechanics \cite{lemaitre2005, lemaitre2012} provides an effective approach for characterizing the fatigue damage evolution and enables prediction of life of components due to various conditions, such as initial damage or surface treatments like shot peening or laser peening. Despite recent development in fatigue models \blue{\cite{mashayekhi2013, yu2016}}, in many cases fatigue life predicted by computational models is orders of magnitude different from the test or field observation. Such differentiation stems from the probabilistic nature of fatigue along with uncertainties that are introduced in geometry, loading, boundary condition, material properties, and modeling itself. One source of uncertainty is the dependency of the initiation and growth of a fatigue crack on the randomness of mechanical properties and material's microstructure \cite{ mcdowell2010, dingreville2010, bolotin2001}.
Moreover, large number of tests are required to establish full spectrum of fatigue strength of a material. Acquiring such experimental data is costly and might not be always accessible. Furthermore, the available experimental measurements might be contaminated with noise introduced by experimental devices or material variabilities. On the other hand, continuum damage mechanics computational models are intrinsically based on simplifying assumption of physical reality and incapable to entirely characterizing the complex fatigue phenomena.
Hence, addressing such uncertainties is crucial for the fatigue life prediction at the design stage that enables reliability and risk assessment of predicted lifetime and open the potential for reduced maintenance and replacement costs of modern engine components, enhanced safety, and increased performance.
}

\blue{
Large number of investigators have addressed the stochastic fatigue phenomena in the literature  \cite{lardner1967, birnbaum1958}. Such efforts involve developing probabilistic techniques to model the fatigue crack propagation under constant or random loading \cite{ortiz1988, yang1983} as well as studying the randomness of the fatigue damage accumulation process caused by random distribution of stress amplitude, number of cycles at a stress amplitude, and fatigue resistance of material \blue{\cite{shen2000, correia2017, zhu2016,zhu2015}}.
Naderi et al.  \cite{naderi2013} simulated fatigue damage and life scatter of metallic components using continuum elasto-plastic damage model.  The progressive fatigue failure was simulated by applying random properties to the finite element domain. Bahloul et al. \cite {bahloul2016probabilistic} used a probabilistic approach to predict fatigue life of cracked structure repaired by high interference fir bushing. In the context of a probabilistic approach, Zhu et al. \cite{zhu2017probabilistic} studied the influence of material variability on the multiaxial low cycle fatigue of notched components by using the Chaboche plasticity model and Fatemi-Socie criterion. Kwon et al. \cite{kwon2011probabilistic} estimated the fatigue life below the constant amplitude fatigue threshold of steel bridges using a probabilistic method on the basis of a bilinear stress life approach. \blue{Zhu et al. \cite{zhu2013bayesian, zhu2012} developed a Bayesian framework for probabilistic fatigue life prediction and uncertainty modeling of aircraft turbine disk alloys. They also quantified the uncertainty of material properties and model uncertainty resulting from choices of different deterministic models in a low cyclic fatigue regime.}
Recently, Babuska et al. \cite{babuvska2016fatigue} proposed a systematic approach for model calibration, model selection and model ranking with reference to stress-life data drawn from fatigue experiments performed on 7075-T6 aluminum alloys. The stress-life models are fitted to the data by maximum likelihood methods and life distribution of the alloy specimen are estimated. The models are also compared and ranked by classical measures of fit based on information criteria. Moreover, Bayesian approach is considered that provides posterior distributions of the model parameter.
}


\blue{
The present study develops a computationally feasible stochastic methodology for prediction and design of material components experiencing cyclic fatigue. 
In this regard, we select a non-linear continuum damage theory for fatigue analysis with Ti6Al4V as the subject material of interes.
Global sensitivity analysis is conducted to determine important model parameters influencing fatigue life prediction. To provide information for model calibration, fatigue tests are performed on coupons of Ti6Al4V material at different stress levels. In order to factor the inherent uncertainties in data and model inadequacy into a statistical analysis for the calibration of the system, a Bayesian framework is also developed for defining, updating, and quantifying uncertainties in the model, the experimental data, and the target quantities of interest. The calibrated model is then utilized for computational life prediction of hardware under uncertainty. In this regard, an illustrative inverse design scenario is taken into account in which surface treatment is considered as a design variable. The optimal design variable and its reliability for a target performance (lifetime) expected from the component is assessed.
}

\blue{
This paper is structured as follows. A summary of the fatigue damage model and laser peening modeling is presented in Section 2. Section 3 describes the computational methods employed for uncertainty treatment. This involves global sensitivity analysis that is performed on the model as well as Bayesian framework for statistical calibration of the model against experimental data with quantified uncertainties. In Section 4, the results of sensitivity analysis, the experimental results used in the statistical analyses, and statistical calibration of fatigue model are presented. Computational prediction, design decision making under uncertainty, along with an illustrative design example are also discussed in this section. A summary and concluding remarks are given in Section 5.
}

\section{Fatigue Analysis Using Continuum Damage Mechanics}\label{sec:damage}

In the continuum damage mechanics (CDM) approach, the mechanical behavior of damaged material is described by a set of constitutive and damage evolution equations. Plasticity and damage are tightly coupled in the CDM formulation. At any instant for a given load increment, plasticity increment depends on the current damage state and in turn, the damage increment depends on the current plasticity state. For the case of high cycle fatigue,  damage is always very localized and occurs at a scale (micro scale) much smaller than the scale of Representative Volume Element (meso scale), where the state of the material remains essentially elastic. 
The damage in metal often initiates in localized grains and grain boundaries or at inclusions where stress concentration builds up to cause plastic flow at a few localized grains, thus plasticity and damage occurs in a very localized region inside the Representative Volume Element (RVE), but there is no overall plasticity at the RVE scale. To account for the plasticity and damage interaction in the CDM model for the case of high cycle fatigue loading, a two-scale model first proposed by Lemaitre \cite{lemaitre1999two} is followed in this paper.

At the RVE scale or mesoscale, the stress is denoted as $\sigma_{ij}$, the total strain as $\epsilon_{ij}$, elastic strain as $\epsilon^{e}_{ij}$ and plastic strain as $\epsilon^{p}_{ij}$. The values at the microscale are denoted by an upper script $\mu$.
 A finite element analysis is performed to obtain the stress $\sigma_{ij}$ at a location of interest due to remote loading, and the corresponding total strain $\epsilon_{ij}$ and elastic strain $\epsilon^{e}_{ij}$ are computed. For high cycle fatigue loading, the plastic strain $\epsilon^{p}_{ij}$ is normally zero due to remote loading. If the material is subjected to laser or shot peening, then the resulting plastic strain $\epsilon^{p}_{ij}$ is non-zero at the meso scale. Next, a scale transition law based on the localization laws of Eshelby and Kroner is utilized to derive the corresponding micro elastic strain $\epsilon^{\mu e}_{ij}$ and micro plastic strain $\epsilon^{\mu p}_{ij}$ from the strains at the meso level.
The Eshelby-Kroner localization laws \cite{Eshelby, Kroner}  are written as,
\begin{eqnarray}
\epsilon^{\mu D}_{ij} & = & \frac{1}{1-bD}
\left[\epsilon^{D}_{ij} + b\left((1-D)\epsilon^{\mu p}_{ij}-\epsilon^{p}_{ij}\right)\right], \nonumber \\
\epsilon^{\mu}_{H} & = & \frac{1}{1-aD}\epsilon_{H}, \label{eq:eshelby}
\end{eqnarray}
the superscript $D$ refers to the deviatoric part and the subscipt $H$ to the hydrostatic part. The constants in the localization laws are given as,
\begin{eqnarray*}
a & = & \frac{1+\nu}{3(1-\nu)}, \\
b & = & \frac{2}{15}\frac{4-5\nu}{1-\nu}.
\end{eqnarray*}
Once the corresponding micro strains are known from the meso strain values at the RVE using the localization laws, the following set of constitutive and evolution laws are used at the micro scale. A crack is assumed to have initiated when the damage at the micro level reaches a critical value,
\begin{eqnarray}
\epsilon^{\mu}_{ij} & = &
\epsilon^{\mu e}_{ij} + \epsilon^{\mu p}_{ij}, \nonumber \\
\epsilon^{\mu e}_{ij} & = &
\frac{1 + \nu}{E}\hat{\sigma}^{\mu}_{ij}
-\frac{\nu}{E}\hat{\sigma}^{\mu}_{kk}\delta_{ij}, \nonumber \\
\hat{\sigma}^{\mu}_{ij} & = & \frac{\sigma}{1-D}, \nonumber \\
\dot{\epsilon}^{\mu p}_{ij} & = &
\frac{3}{2}\frac{\hat{\sigma}^{\mu D}_{ij} -
\chi^{\mu}_{ij}}
{\left(\hat{\bm{\sigma}}^{\mu D} - \bm{\chi}^{\mu}\right)_{eq}}
\dot{p}^{\mu},  \label{eq:main} \\
\dot{\chi}^{\mu}_{ij} & = & \frac{2}{3}C_{y}\dot{\epsilon}^{\mu p}_{ij}(1-D), \nonumber \\
\dot{D} & = & \left(\frac{Y^{\mu}}{S}\right)^{s}, \nonumber
\end{eqnarray}
where $\hat{\sigma}^{\mu}_{ij}$ is the effective stress at microscale, $\chi^{\mu}_{ij}$ is the kinematic hardening tensor at the microscale, $C_{y}$ is the plastic modulus, and damage strength $S$, damage exponent $s$ and critical damage $D_{c}$ are material parameters. The quantity $Y$ is the energy release rate term and is given in terms of the stress components as
\begin{equation}
Y^{\mu}_{n+1} =
\frac{1+\nu}{E}
\left(
\hat{\bm{\sigma}}^{\mu}_{n+1}
\mbox{:}
\hat{\bm{\sigma}}^{\mu}_{n+1}
\right)
-
\frac{\nu}{2E}
\left(
\mbox{tr }
\hat{\bm{\sigma}}^{\mu}_{n+1}
\right)
^{2}.
\end{equation}

The local stress history $\sigma_{ij}$ is  obtained by determining the local stress at the point of interest due to a reference remote load, from finite element computation. Next,  for a given remote load history, the local stresses are  scaled appropriately, and the local stress history is obtained. The residual strain $\epsilon^{p}_{ij}$ (due to laser peening or other technique) is a constant value obtained from a separate analysis. Each load cycle is broken up into a number of load increments and meso strain increment is obtained as $\Delta \boldsymbol{\epsilon} = \boldsymbol{\epsilon}_{n+1} - \boldsymbol{\epsilon}_{n}$. The incremental micro strains are computed by assuming they are completely elastic from the meso strain increments. If the new state remains within the yield surface, then the assumption of purely elastic micro strains is acceptable and we move to the next load increment. On the other hand, if there is plastic yielding, then the incremental micro  plastic strain is computed and the incremental micro elastic strain is corrected so that the material state returns to the yield surface. The incremental damage is calculated from incremental micro plastic strain and added to the  accumulated damage and the process continues until the total damage crosses a threshold value which indicates initiation of a crack. The number of cycles for the damage to reach the critical value is the fatigue crack initiation  life of the component under high cyclic fatigue loading. The details of obtaining incremental micro plastic strain and incremental micro damage can be found in \cite{lemaitre1999two}.

\section{Uncertainty Treatments and Probabilistic Design}\label{sec:uncertainty}

The treatment of uncertainty in predictive modeling involves three distinct processes:
1)~the statistical inverse process, in which probability densities of random model parameters and modeling errors in the theoretical structure of computational model are estimated using measurement data
2)~the statistical forward process involving propagation of input uncertainties through the computational model to quantify the uncertainties in quantity of interests (QoI's) and
3)~ultimately using the stochastic model with quantified uncertainties due to both modeling and measurement errors for control or design decision making under uncertainty.

In this section, we describe the methods and techniques we have employed for uncertainty treatment in cyclic fatigue life prediction. These involve propagation of uncertainty through the forward model and solution process using Monte Carlo methods, global parameter sensitivity analysis, and Bayesian methods for model calibration with quantified uncertainty.

\subsection{Forward Uncertainty Propagation and Sensitivity Analysis}

In order to illustrate our proposed probabilistic design, we use the notion of \textit{black-box model}. In this setting, the continuum damage constitutive relation presented in previous section along with boundary and initial conditions are cast into an abstract form of the computational (forward) model. 
\blue{In a black-box (input-output) model of a system, the underlying character or physics of the relations involved as well as numerical solutions are hidden. This notion is used in this section to lay down a general framework of uncertainty quantification in computational models. It is shown later that non-intrusive probabilistic methods require only the input and output of the model for forward and inverse uncertainty assessment. In this regard, the probabilistic design method presented in this section is general and can be implemented for any material model.
}

\begin{figure}[h]
\begin{center}
\includegraphics[trim = 0mm 85mm 0mm 45mm, clip,width=0.7\textwidth]{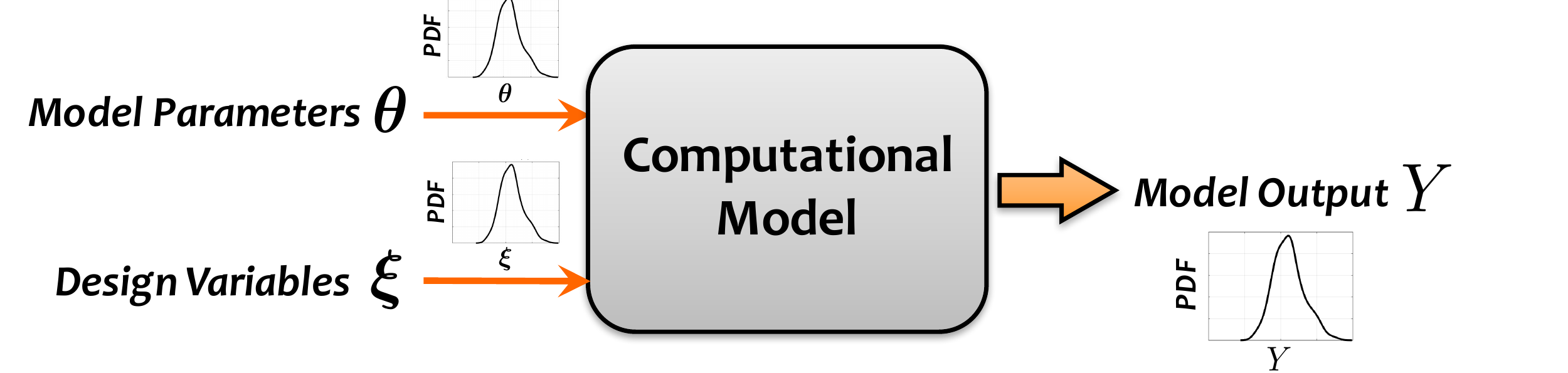}
\end{center}
\vspace{-3mm}
\caption{
Computational model as a black-box (input-output). \blue{The underlying physics of the fatigue damage model is hidden in order to develop a general uncertainty treatment process.}
}
\label{fig:forward}
\end{figure}

As shown in Figure \ref{fig:forward}, the inputs of a computational model can be categorized to {model parameters} (vector of $\boldsymbol{\theta}$) and {design variables} (vector of $\boldsymbol{\xi}$). Model parameters are the inputs that cannot be controlled while design variables are the ones that can be controlled and changed to improve the design. Given these inputs, the computational model evaluates the output or response variable $Y(\boldsymbol{\theta},\boldsymbol{\xi})$. The goal of constructing and solving the computational model is to compute specific quantities of interest (QoIs) that can be determined from model output $Y$.

For the cyclic fatigue life prediction, the model parameters can be geometry of component and material parameters of the continuum damage model (indicating strength, microstructure, in-service damage) and the design variables are residual stress (can be controlled through surface treatments), operational load, and usage cycles. Output response in this case is the damage variable that will be used to assess and predict the life of the hardware (QoI) due to fatigue damage.

Presence of inevitable uncertainties in modeling and experimental data results in deviation of computational prediction from physical reality. The main source of uncertainties is variability of input parameters such as manufacturing imprecision in geometry, variation of material properties, noises in experimental measurements used for damage model calibration and validation \cite{prudencio2013, prudencio2014ijnme, prudencio2014compB}. The variability of these factors cannot be reduced in an industrial setting although it causes response variation and may result in catastrophic events when the product fails. Accounting for such uncertainties in fatigue life prediction involves two main challenges: (1) modeling the effect of uncertainty in input parameters (2) assessing the effect of uncertainty on life prediction of the hardware.
In order to cope with these uncertainties, we make use of probability theory. In this setting, model parameters and design variables are not deterministic; they are random variables characterized by probability density functions (PDFs). Therefore the forward model is a stochastic problem in which having $\boldsymbol{\theta}$ and $\boldsymbol{\xi}$ as PDFs, the response variable $Y$ will become a random variable.
Second challenge involves developing techniques to propagate these uncertainties through models and provide statistical solution for life prediction. Commonly used methods to propagate these random variables through
models include the large family of Monte Carlo methods. Monte Carlo methods lead to a collection of independent realisations of the forward model to solve. One of the main reasons that
these methods are so popular in practice is that they are completely non-intrusive, requiring only
an existing deterministic solver. The convergence of the basic Monte Carlo method is guaranteed
under very weak assumptions and independent of the dimension of the parameter space. Probability distribution of $Y$ obtained from Monte Carlo simulation indicates the sensitivity of response to the variation of inputs and level of confidence of the computational prediction.

The core computational kernel of stochastic algorithms based on the Monte Carlo is the repeated
evaluation of the forward model with different parameters. In the context of fatigue damage life
prediction models that is described by a system of partial differential equations, it is usually the solution of the deterministic forward model, the map between a single realisation of the stochastic parameter and a quantity of interest in the output of a model that dominates the overall cost of solving the statistical problem.


In the context of forward propagation of uncertainty, parameter sensitivity analysis can be considered as one of the most useful techniques in probabilistic design. It is the study of how the change in input factors of a model, qualitatively and quantitatively, affects the variation and uncertainty in the model outputs.
As shown in the next section, such analysis determines the most effective model parameters and reduce computational cost of statistical calibration and aids in robust design of the system.
A qualitative way to visualize the sensitivity of parameters is to produce scatter plots
\cite{saltelli2008} which are clouds of points indicating the model output versus each parameter,
constructed by random variations in all parameters. 
\blue{
The parameters with significant contribution to the output are the ones with distinct pattern in the scatter-plot cloud.
}
Among different quantitative techniques for sensitivity analysis of model output, variance-based methods \cite{Cukier1973,Sobol1990,Sobol1993} have been proven to be effective and well suited for practical applications.
In these methods the sensitivity of the output to an input variable is measured by the amount of (conditional) variance in the output caused by that specific input.

\subsubsection{Variance-based sensitivity analysis}

Consider a model $Y = f (\theta_1, \theta_2, \dots , \theta_k)$ with $k$ uncertain input factors $\theta_i$, in which $Y$ is model output and $f$ is a square integrable function.
Using the Hoeffding decomposition of $f$ \cite{Sobol1990, Sobol1993} and conditional expectations of the model, $E(Y|\theta_i)$,
one can derive the following decomposition for the output variance, $V(Y)$,
\begin{equation}\label{eq:decomp_V}
V(Y) = \sum_{i} V_i + \sum_{i} \sum_{j>i} V_{ij} + \dots + V_{12\dots k},
\end{equation}
where
\begin{equation}\label{eq:variances}
\begin{matrix}
V_i    =& V \left( f_i(\theta_i) \right)        =& V_{\theta_i} \left(E_{\boldsymbol{\theta}_{\sim i}} (Y|\theta_i)\right), & \\
V_{ij} =& V \left( f_{ij}(\theta_i,\theta_j) \right) =&  V_{\theta_i \theta_j} \left(E_{\boldsymbol{\theta}_{\sim {ij}}} (Y|\theta_i,\theta_j)\right)  & - V_{\theta_i} \left(E_{\boldsymbol{\theta}_{\sim i}} (Y|\theta_i)\right) \\
       & &  - V_{\theta_j} \left(E_{\boldsymbol{\theta}_{\sim j}} (Y|\theta_j)\right). &
\end{matrix}
\end{equation}
In (\ref{eq:variances}) $\theta_i$ is the $i$-th factor, $\boldsymbol{\theta}_{\sim i}$ indicates the matrix of all factors except $\theta_i$, $V_{\theta_i} \left(E_{\boldsymbol{\theta}_{\sim i}} (Y|\theta_i)\right)$ and  $V_{\theta_j} \left(E_{\boldsymbol{\theta}_{\sim j}} (Y|\theta_j)\right)$ are called the first-order effects, and $V_{\theta_i \theta_j} \left(E_{\boldsymbol{\theta}_{\sim {ij}}} (Y|\theta_i,\theta_j)\right)$ is the joint effect of $(\theta_i;\theta_j)$ on $Y$.
In $V_{\theta_i} \left(E_{\bold{\theta}_{\sim i}} (Y|\theta_i)\right)$, the internal expectation represents the mean of $Y$ taken over all possible values of $\boldsymbol{\theta}_{\sim i}$ while $\theta_i$ is fixed, and the outer variance is taken over all possible values of $\theta_i$.
Therefore, $V_{\theta_i} \left(E_{\boldsymbol{\theta}_{\sim i}} (Y|\theta_i)\right)$ is the expected variance reduction when $\theta_i$ is fixed.

For non-additive models, such as the fatigue damage model under study, the model output depends on the interaction among $\theta_{i}$ and $\theta_{j}$.
Decomposing the variance by conditioning with respect to all factors but $\theta_{i}$ leads to,
\begin{equation}
V (Y ) = V_{\boldsymbol{\theta}_{\sim i}} \left(E_{\theta_i} (Y | \boldsymbol{\theta}_{\sim i}) \right) + E_{\boldsymbol{\theta}_{\sim i}} \left(V_{\theta_i} (Y | \boldsymbol{\theta}_{\sim i}) \right),
\end{equation}
and dividing both sides by $V(Y)$, results in a sensitivity measure, so-called the \textit{total effect index} \cite{HommaSaltelli1996, SaltelliTarantola2002},
\begin{equation}\label{eq:ST}
\mathcal{S}_{T_i} = \frac{E_{\boldsymbol{\theta}_{\sim i}} \left(V_{\theta_i} (Y | \boldsymbol{\theta}_{\sim i}) \right)}{V(Y)} = 1- \frac{V_{\boldsymbol{\theta}_{\sim i}} \left(E_{\theta_i} (Y | \boldsymbol{\theta}_{\sim i}) \right)}{V(Y)}.
\end{equation}

In (\ref{eq:ST}), $E_{\boldsymbol{\theta}_{\sim i}} \left(V_{X_i} (Y | \boldsymbol{\theta}_{\sim i}) \right)$ is the residue variance of $Y$ for fixed $\theta_i$ and $V_{\boldsymbol{\theta}_{\sim i}} \left(E_{\theta_i} (Y | \boldsymbol{\theta}_{\sim i}) \right)$ denotes the expected reduction in variance if all values other than $\theta_{i}$ are fixed.
Total effect $\mathcal{S}_{T_i}$ measures the contribution of factor $\theta_i$ to the output variation.
Small values of the total effect index imply that $\theta_{i}$ can be fixed at any value within its range of variability (uncertainty) without appreciably affecting the output.
%

\subsection{Statistical Inverse Theory}

Predictive computational modeling and simulation-based design requires experimental data to be integrated into computational
models in order to calibrate the models and assess their validity.
However, the measurement data (e.g., fatigue test) involves material variability and might be incomplete and contaminated with noise, i.e., error in data.
Moreover, the computational model (e.g., continuum damage model) is constructed based on modeling assumptions and is an imperfect characterizations of reality, i.e., error in model.
The problem of overriding importance is to characterize in a worthwhile way all of these uncertainties, to track their propagation through the solution, and to ultimately determine and quantify the uncertainty in the target QoIs.
Deterministic inverse methods for model calibration, based on optimization techniques, are limited in characterizing the uncertainties in data and computational model.
In this regard we employ Bayesian approaches, based on concurrent treatments of statistical inverse analysis, such as the ones described in~\cite{tarantola,somersalo1,Farrell2015, odenbauskafaghihi, farrell2014}.
The main hypothesis of this theory is that of subjective probability; the parameters $\boldsymbol{\theta}$, the observational data $\mathbf{D}$, and the theoretical model are not deterministic; they are random variables or processes characterized by probability density functions (PDF's).

The calibration process enables one to identify model parameters based on calibration data $\mathbf{D}$ obtained from a set of experiments. Within the context of fatigue life prediction, the calibration process would refer to the identification of the damage model parameters from sets of fatigue test. The solution to the statistical calibration problem, following Bayesian approach,  is the posterior PDF $\pi_{\rm post}(\boldsymbol{\theta}|\mathbf{D})$ for the model parameters, updated from a prior PDF $\pi_{\rm prior}(\boldsymbol{\theta})$  as (see e.g.~\cite{kaipio2006}),
\begin{equation}\label{eq:bayes}
\pi_{\rm post}(\boldsymbol{\theta}|\mathbf{D}) =
\frac{\pi_{\rm like}(\mathbf{D}|\boldsymbol{\theta})\cdot \pi_{\rm prior}(\boldsymbol{\theta})}{\pi_{\rm data}(\mathbf{D})},
\end{equation}
where $\pi_{\rm like}(\mathbf{D}|\boldsymbol{\theta})$ denotes the likelihood PDF and the term,
\begin{equation}
\pi_{\rm data}(\mathbf{D}) = \int \pi_{\rm like}(\mathbf{D}|\boldsymbol{\theta})\cdot \pi_{\rm prior}(\boldsymbol{\theta}) d \boldsymbol{\theta},
\end{equation}
is the normalization value that makes (\ref{eq:bayes}) a PDF.

Prior PDF $\pi_{\rm prior}$ represents initial knowledge or information we may have about parameters $\boldsymbol{\theta}$ before observing the data.
In structural mechanics and from engineering perspective one might simply assume an appropriate range of parameters from prior experiments and analyses on similar material and system.
Likelihood PDF $\pi_{\rm like}(\mathbf{D}|\boldsymbol{\theta})$ encapsulates assumptions about the discrepancy
between the values $\mathbf{D}$ that are measured and the values that can be computed with the computational model $\mathbf{Y}(\boldsymbol{\theta})$.
One common approach to defining a likelihood function is based on additive total error (data noise and model
inadequacy) hypothesis. Assuming total error being Gaussian random variables of zero mean, the form of likelihood can be postulated as
\begin{equation}\label{eq:likelihood}
\pi_{\rm like}(\mathbf{D}| \boldsymbol{\theta}) \propto \exp
\left\lbrace
-\frac{1}{2} \left[\mathbf{Y}(\boldsymbol{\theta}) - \mathbf{D}\right]^T \mathbf{C}^{-1}
\left[\mathbf{Y}(\boldsymbol{\theta}) - \mathbf{D}\right]
\right\rbrace,
\end{equation}
where
$\mathbf{C}$ is a covariance matrix, $\mathbf{D}$ denotes experimental data, and $\mathbf{Y}(\boldsymbol{\theta})$ is the model output.

\section{Results and Discussions}\label{sec:results}

\begin{figure}[h]
\begin{center}
\includegraphics[trim = 10mm 55mm 40mm 49mm, clip,width=0.7\textwidth]{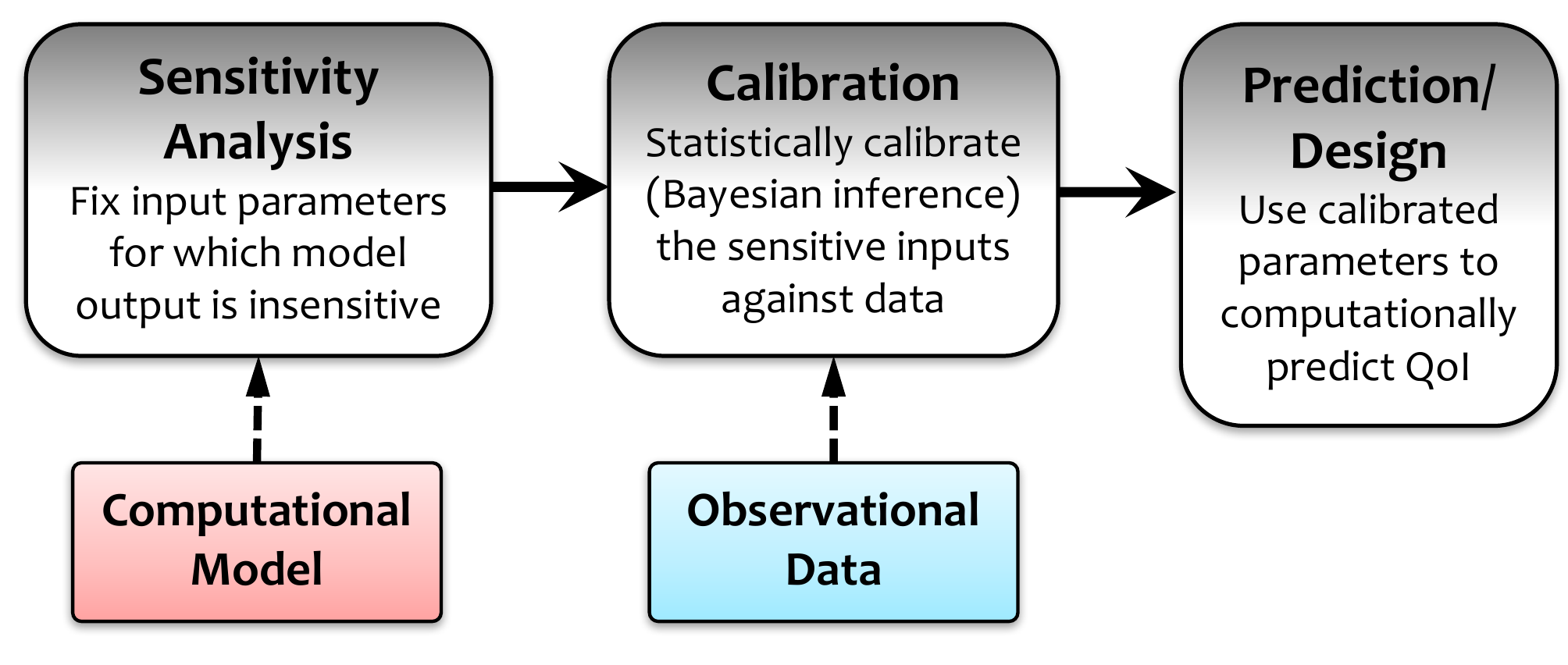}
\end{center}
\vspace{-5mm}
\caption{
\blue{
The proposed framework of uncertainty treatment in computational model and observational data.
First the computational fatigue damage model is subjected to sensitivity analysis to determine significant input parameters of the model.
Next using experimental data the important parameters of model are calibrated. Utilizing statistical inverse methods (Bayesian approach) for model calibration, results in encapsulating all the uncertainties due to fatigue test data and inadequacy of the fatigue damage model into the probability distribution function of the model parameters.
Finally, the statistically calibrated model is used to make prediction or design decision with quantified uncertainty.
}
}
\label{fig:flowchart}
\end{figure}

The uncertainty treatment methods described in previous section is employed to integrate experimental data and computational fatigue damage model in order to assess the reliability of computational prediction and corresponding design decision making. Due to high computational cost of Bayesian inference, model sensitivity analysis is initially performed in order to discard irrelevant parameters from statistical calibration. Bayesian calibration is then conducted to statistically infer the most important parameters of the model using the information provided by fatigue test data. Ultimately, the calibrated model with quantified uncertainty is employed to make predictions about the fatigue strength of the material in the defined loaded stress state. \blue{The process of reliability assessment of the computational model prediction is depicted in Figure \ref{fig:flowchart}.}

%

\subsection{Sensitivity Analysis of Fatigue Damage Model}

\begin{figure}[h!]
\begin{center}
\includegraphics[trim = 0mm 0mm 0mm 0mm, clip,width=0.31\textwidth]{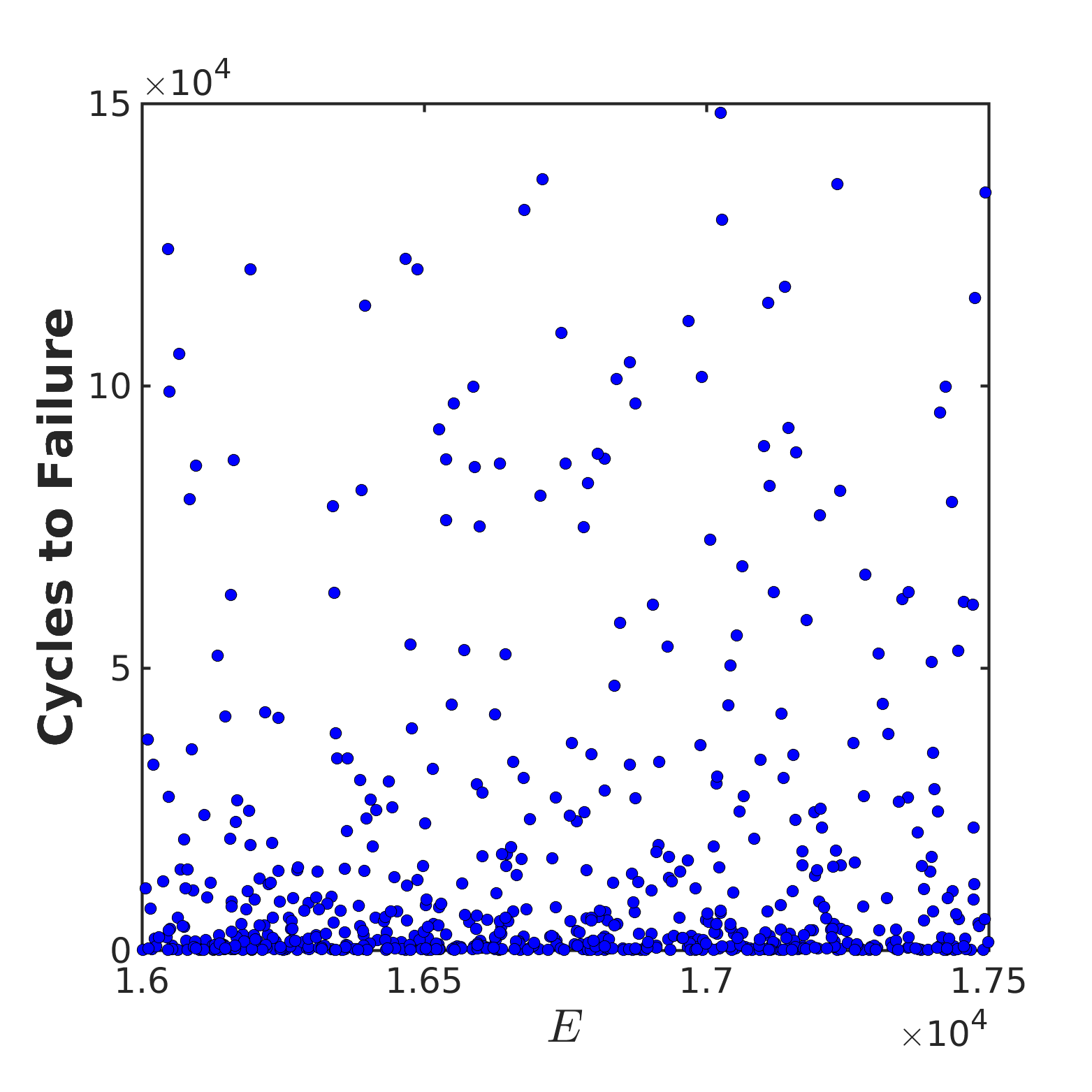}
~
\includegraphics[trim = 0mm 0mm 0mm 0mm, clip,width=0.31\textwidth]{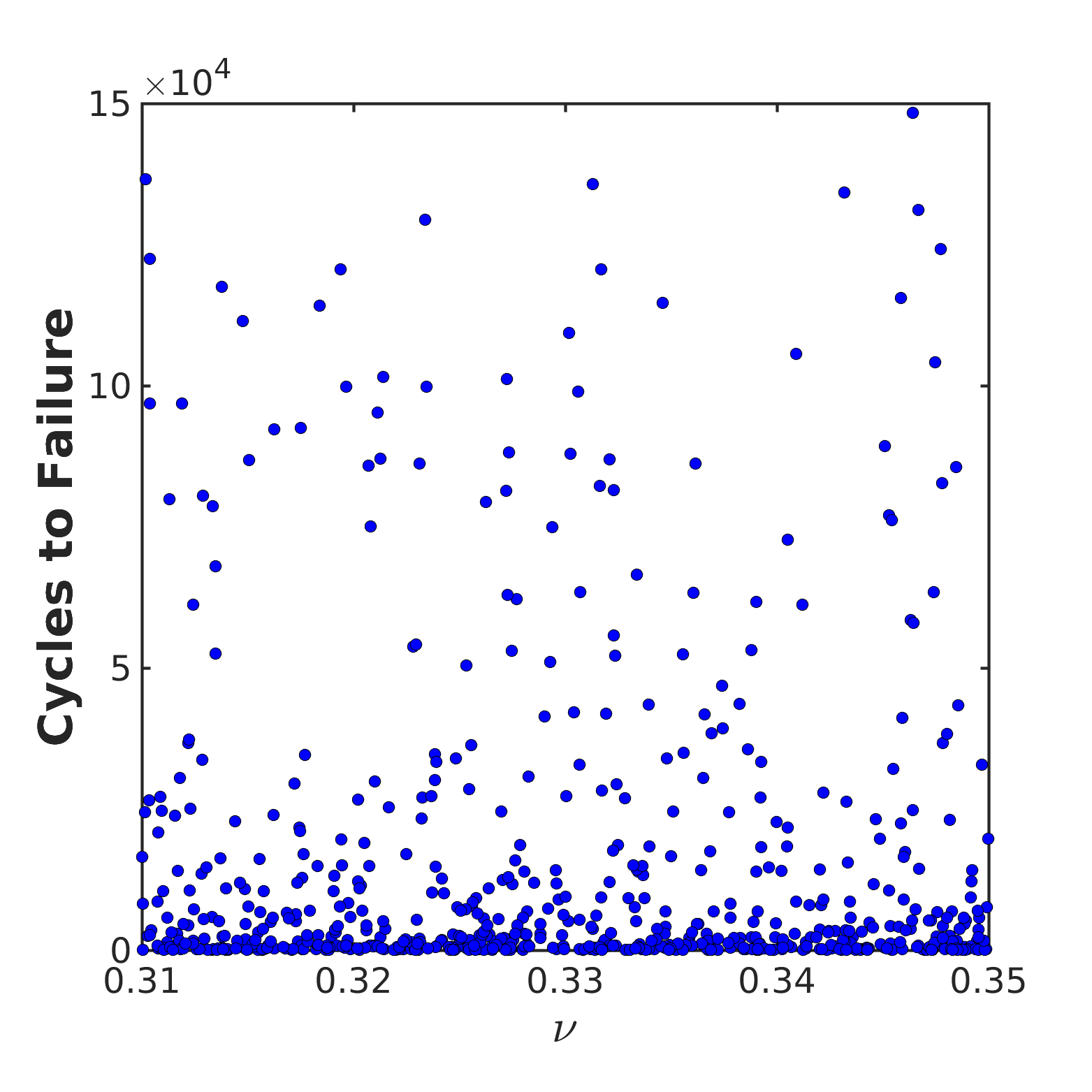}
~
\includegraphics[trim = 0mm 0mm 0mm 0mm, clip,width=0.31\textwidth]{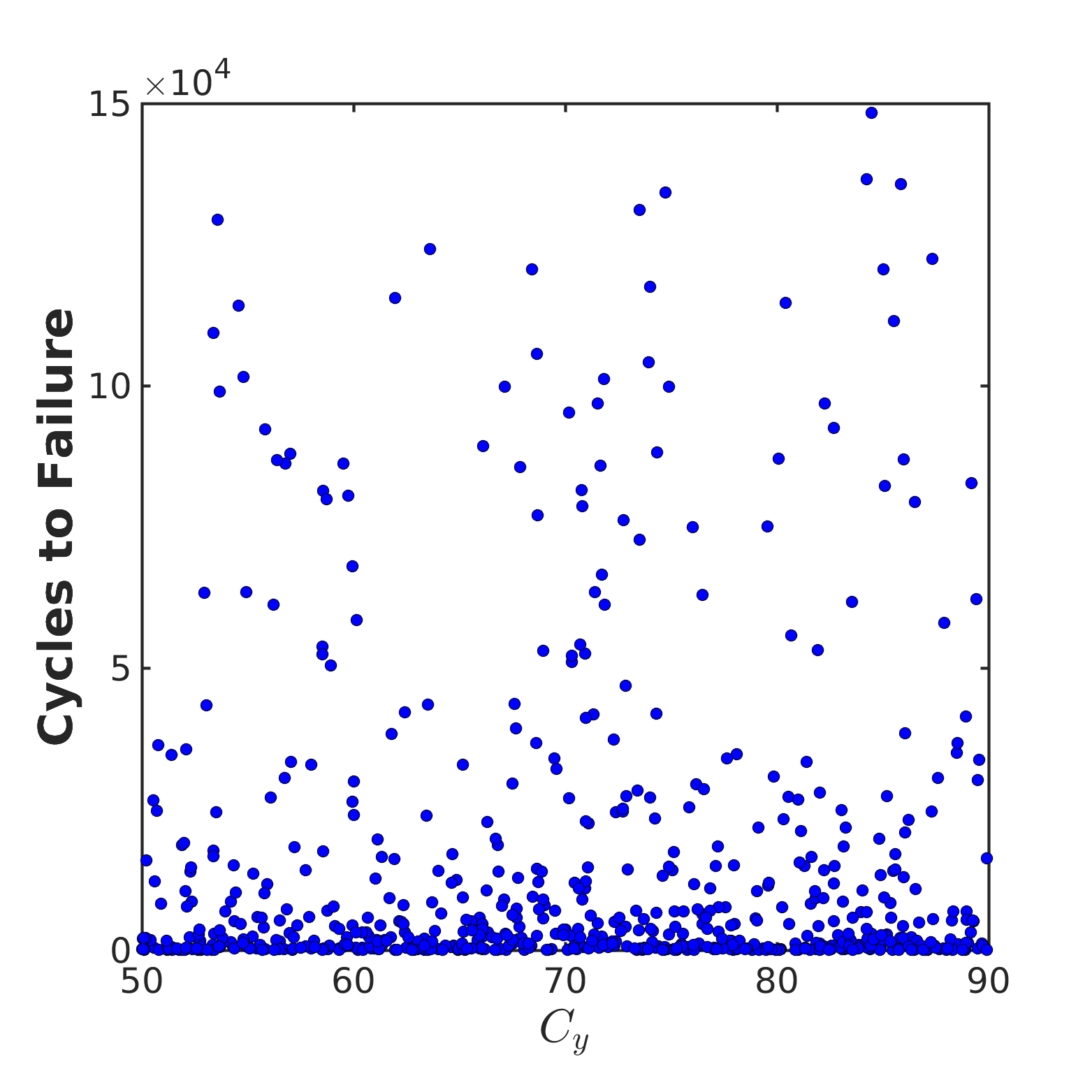}
\\
\includegraphics[trim = 0mm 0mm 0mm 0mm, clip,width=0.31\textwidth]{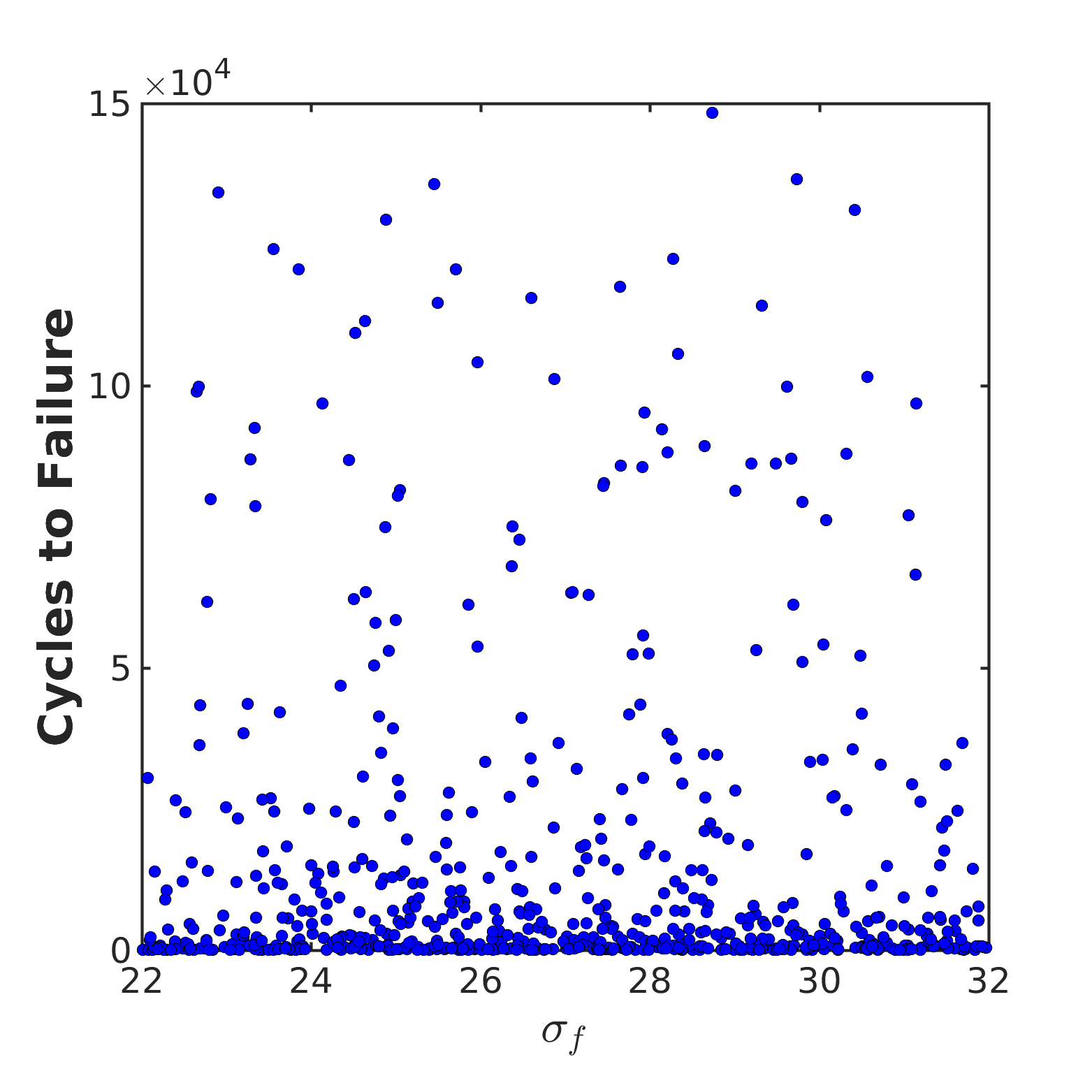}
~
\includegraphics[trim = 0mm 0mm 0mm 0mm, clip,width=0.31\textwidth]{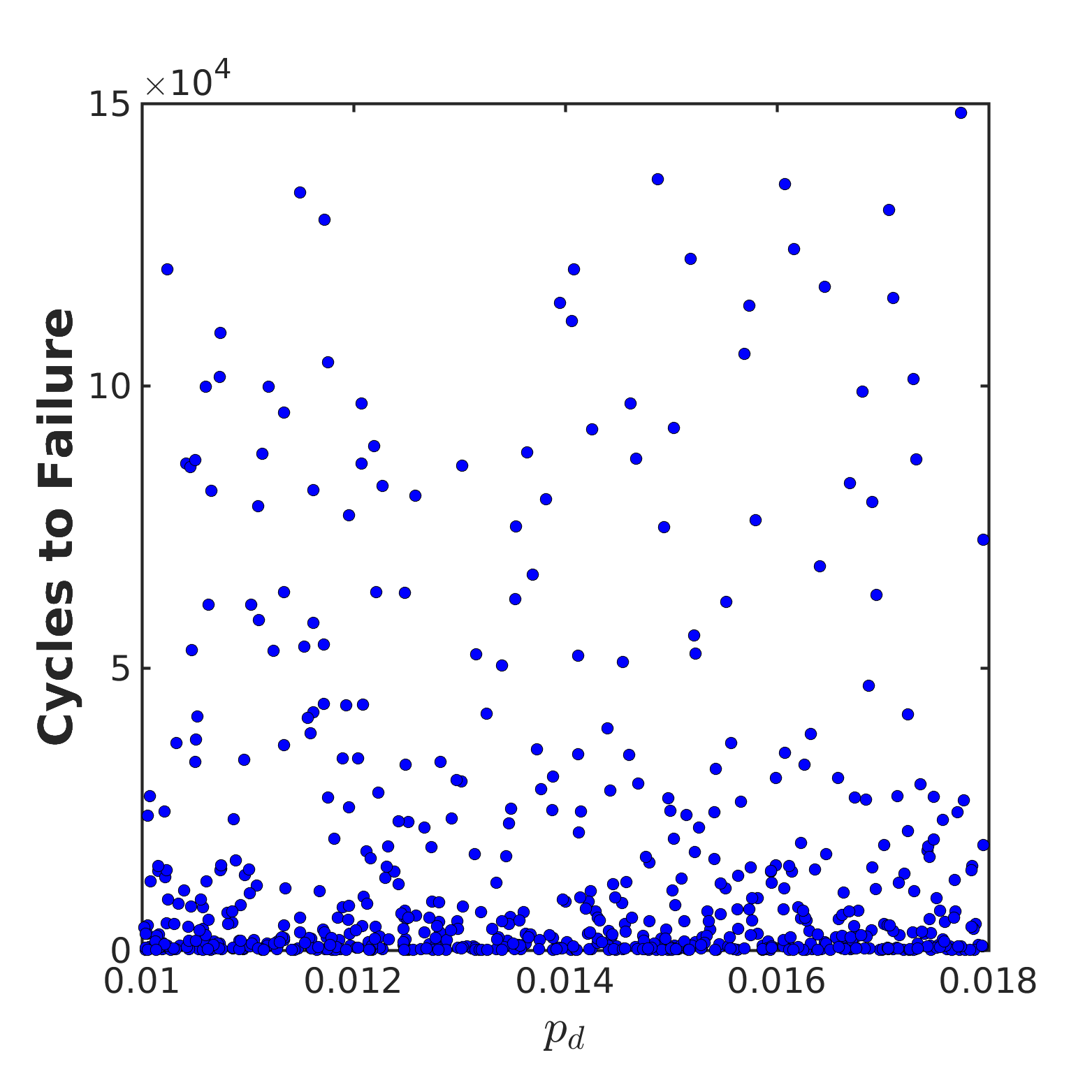}
~
\includegraphics[trim = 0mm 0mm 0mm 0mm, clip,width=0.31\textwidth]{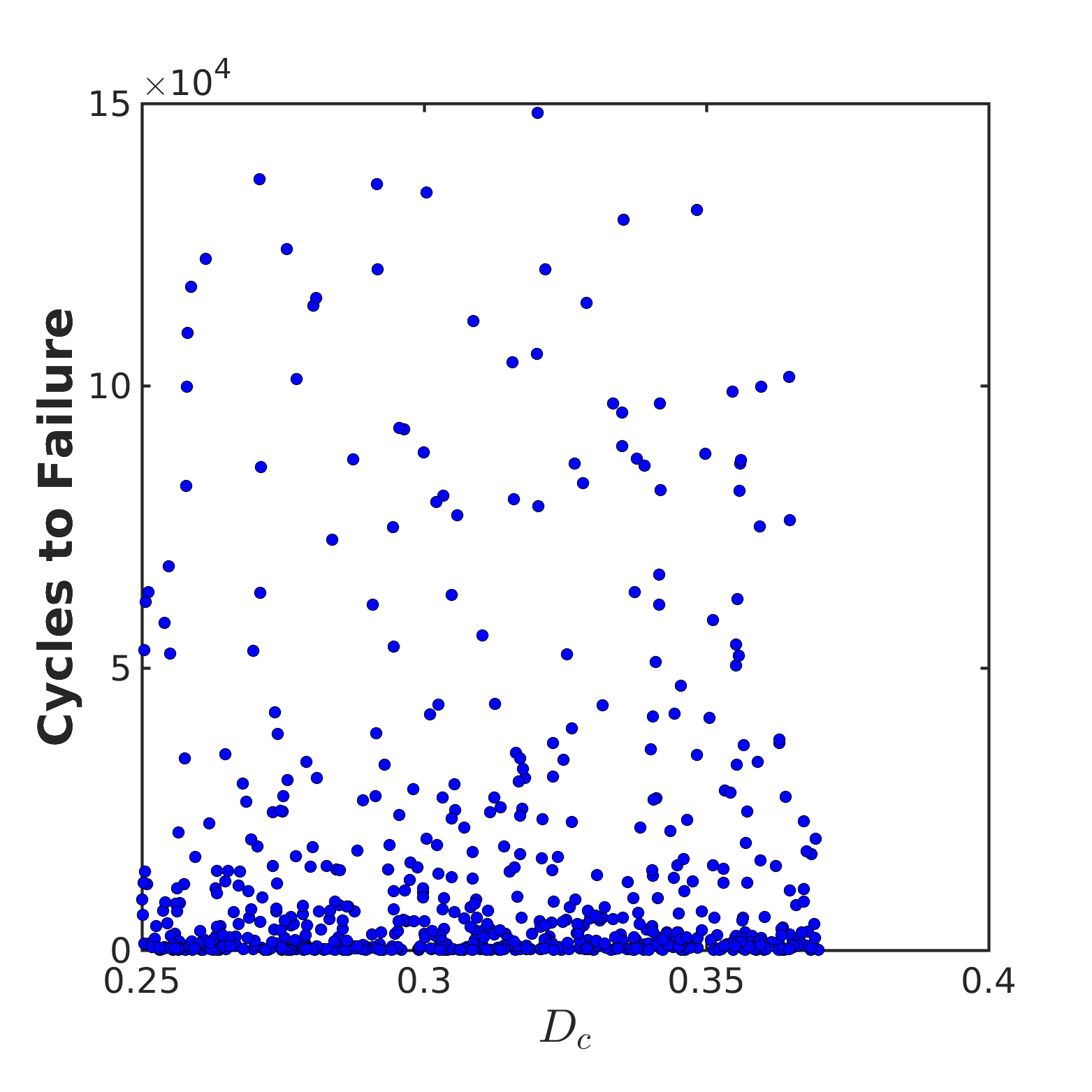}
\end{center}
\vspace{-10mm}
\caption{
Scatterplots illustrating a qualitative sensitivity analysis for the fatigue damage using 1000 Monte Carlo samples of the parameter space.}
\label{fig:scatter1}
\end{figure}

\begin{figure}[h!]
\begin{center}
\includegraphics[trim = 120mm 0mm 120mm 0mm, clip,width=0.48\textwidth]{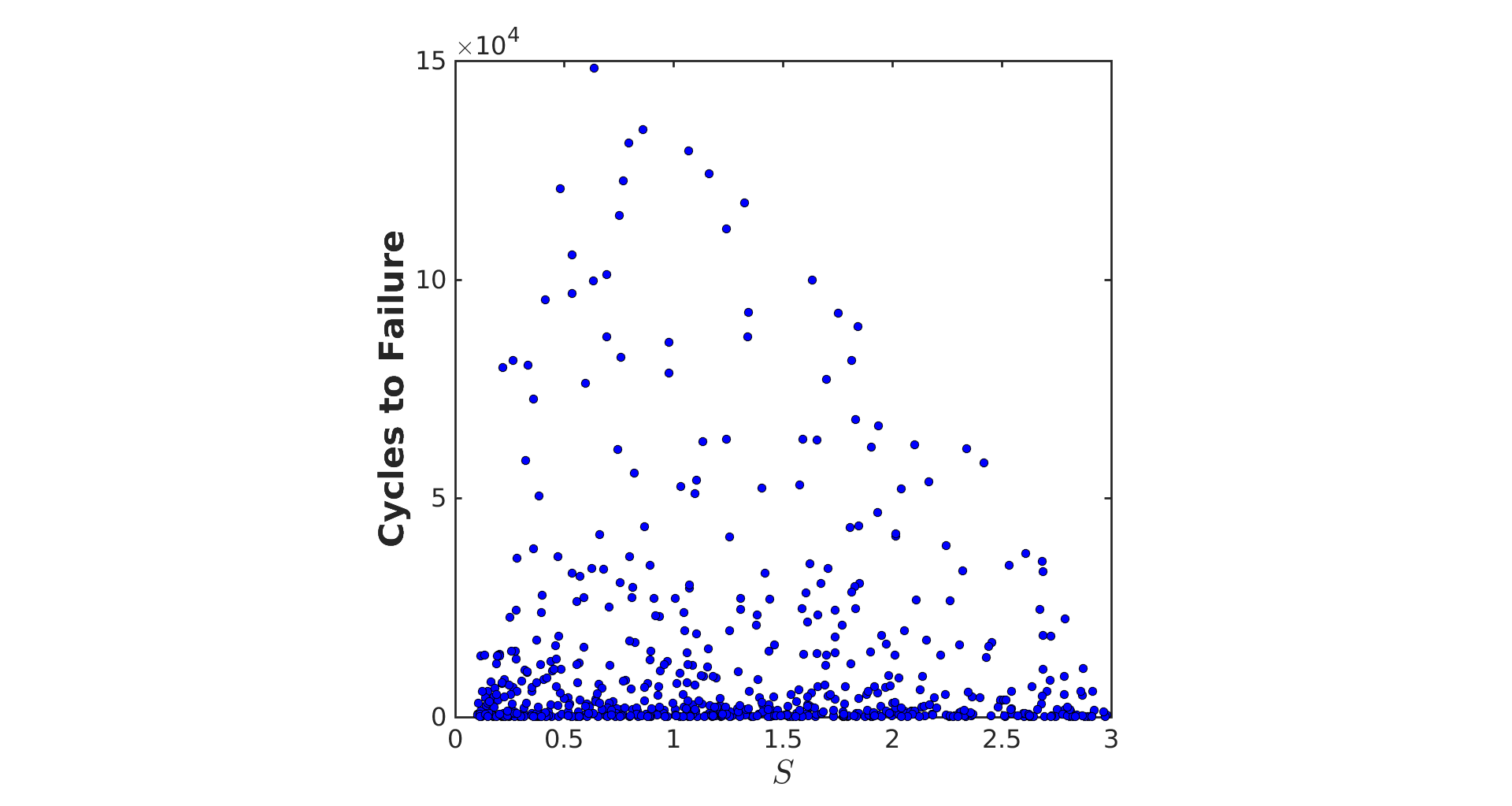}
~
\includegraphics[trim = 120mm 0mm 120mm 0mm, clip,width=0.48\textwidth]{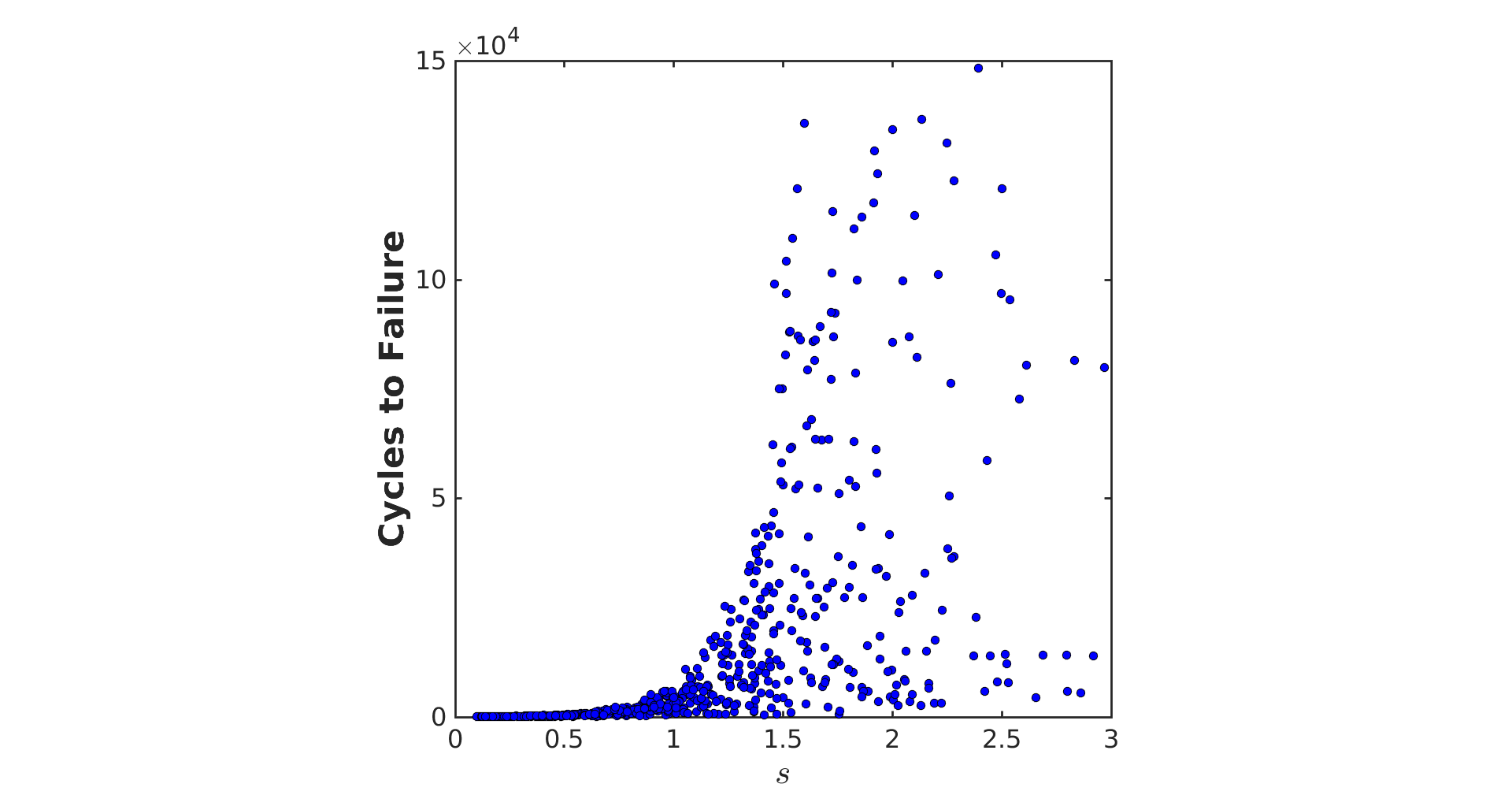}
\end{center}
\vspace{-10mm}
\caption{
Scatterplots illustrating a qualitative sensitivity analysis for the fatigue damage using 1000 Monte Carlo samples of the parameter space.}
\label{fig:scatter2}
\end{figure}

In order to determine the relative level of importance of each parameter with respect to  others, quantitative sensitivity analysis is conducted on fatigue damage model.
For sensitivity scenario, a specimen without surface treatment and under completely reverse stress is taken into account. The amplitude of alternating applied stress is assumed to be 50 ksi and the number of load cycles before failure is considered as model output $Y$.
With respect to this observable, the scatterplots for some of the damage model parameters are shown in Figures \ref{fig:scatter1} and \ref{fig:scatter2}. These plots were obtained by sampling the parameters from given distributions and computing the model output at each sample. The initial distributions of parameters are assumed to be uniform and in the range of reported values for Ti6Al4V material in the literature \cite{allahverdizadeh2012,naderi2013,hammer2012,yatnalkar2010,zherebtsov2005,lemaitre1999two} and are presented in Table \ref{table:param_sen}.

\begin{table}[]
\centering
\begin{tabular}{ll|l}
\hline
Young's modulus          & $E$              & $\mathcal{U}(16000,17500)$ ksi    \\
Poisson ratio                & $\nu$            & $\mathcal{U}(0.31,0.35)$         \\
Plastic modulus            & $C_y$          & $\mathcal{U}(50,90)$ ksi    \\
Ultimate strength          & $\sigma_u$ & $\mathcal{U}(130,145)$ ksi     \\
Yield stress                   & $\sigma_y$ & $\mathcal{U}(80,110)$ ksi     \\
Fatigue limit                  & $\sigma_f$  & $\mathcal{U}(22,32)$ ksi     \\
Damage denominator   & $S$             & $\mathcal{U}(0.1,4)$ \\
Damage exponent        & $s$              & $\mathcal{U}(0.1,4)$ \\
Plastic strain threshold & $p_d$          & $\mathcal{U}(0.01,0.018)$   \\
Critical Damage            & $D_c$         & $\mathcal{U}(0.25,0.37)$   \\ \hline
\end{tabular}
\caption{Distribution of Ti6Al4V material parameters for sensitivity analysis of the fatigue model. $\mathcal{U}(\cdot,\cdot)$ indicates uniform distribution.}
\label{table:param_sen}
\end{table}

%
%


The qualitative sensitivity analysis obtained from scatter plots is then confirmed by computing total sensitivity indices (\ref{eq:ST}).
In the vast majority of applications, it is not trivial to analytically calculate the total-effect sensitivity indices.
Moreover, Monte Carlo method for estimating the conditional variances requires multi-dimensional integrals in the input factors space and it is computationally intractable.
Saltelli \cite{Saltelli2002, saltelli2008, Saltelli2010, HommaSaltelli1996}
proposed a Monte Carlo scheme to determine the indices with a minimized computational cost of estimating multi-dimensional integrals.
Such method is a generalization of the original approach proposed by Sobol' \cite{Sobol1990, Sobol1993} and can be summarized by the following procedure:

\begin{itemize}

\item Create two $N$ independent samples of $k$ parameters and store them in matrices $\bold{A}$ and $\bold{B}$.

\item Generate a matrix $\bold{D}_i$  established by all columns of $\bold{A}$ except the $i$th column,
which is from $\bold{B}$.

\item Evaluate the model outputs for all the matrices
$ \mathcal{Y}_A = f(\textbf{A}) ;
 \mathcal{Y}_B = f(\textbf{B})  ;
  \mathcal{Y}_{D_i} = f(\textbf{D}_i)$.

\item The total-effect indices can then be estimated using following estimator \cite{Saltelli2010},
\begin{equation}\label{ST_estim}
\mathcal{S}_{T_i} \approx
1- \frac{\frac{1}{N} \sum_{j=1}^N \mathcal{Y}_A^{(j)} \mathcal{Y}_{D_i}^{(j)} - \left( \frac{1}{N} \sum_{j=1}^N \mathcal{Y}_A^{(j)} \right)^2}{\frac{1}{N} \sum_{j=1}^N \mathcal{Y}_A^{(j)} \mathcal{Y}_A^{(j)} -
\left( \frac{1}{N} \sum_{j=1}^N \mathcal{Y}_A^{(j)} \right)^2}.
\end{equation}
\end{itemize}

The relation (\ref{ST_estim}) is obtained from (\ref{eq:ST}) by replacing the Monte-Carlo estimators proposed by \cite{Saltelli2002},
\begin{equation}
V_{\boldsymbol{\theta}_{\sim i}}[E_{\theta_i}(Y|\boldsymbol{\theta}_{\sim i})] \approx
\frac{1}{N} \sum_{j=1}^N \mathcal{Y}_A^{(j)} \mathcal{Y}_{D_i}^{(j)} - E^2(Y),
\end{equation}
and
\begin{equation}
V(Y) = E(Y^2) - E^2(Y) =
\frac{1}{N} \sum_{j=1}^N \mathcal{Y}_A^{(j)} \mathcal{Y}_A^{(j)} -
\left( \frac{1}{N} \sum_{j=1}^N \mathcal{Y}_A^{(j)} \right)^2.
\end{equation}

\begin{figure}[h!]
\begin{center}
\includegraphics[trim = 0mm 0mm 0mm 0mm, clip,width=0.6\textwidth]{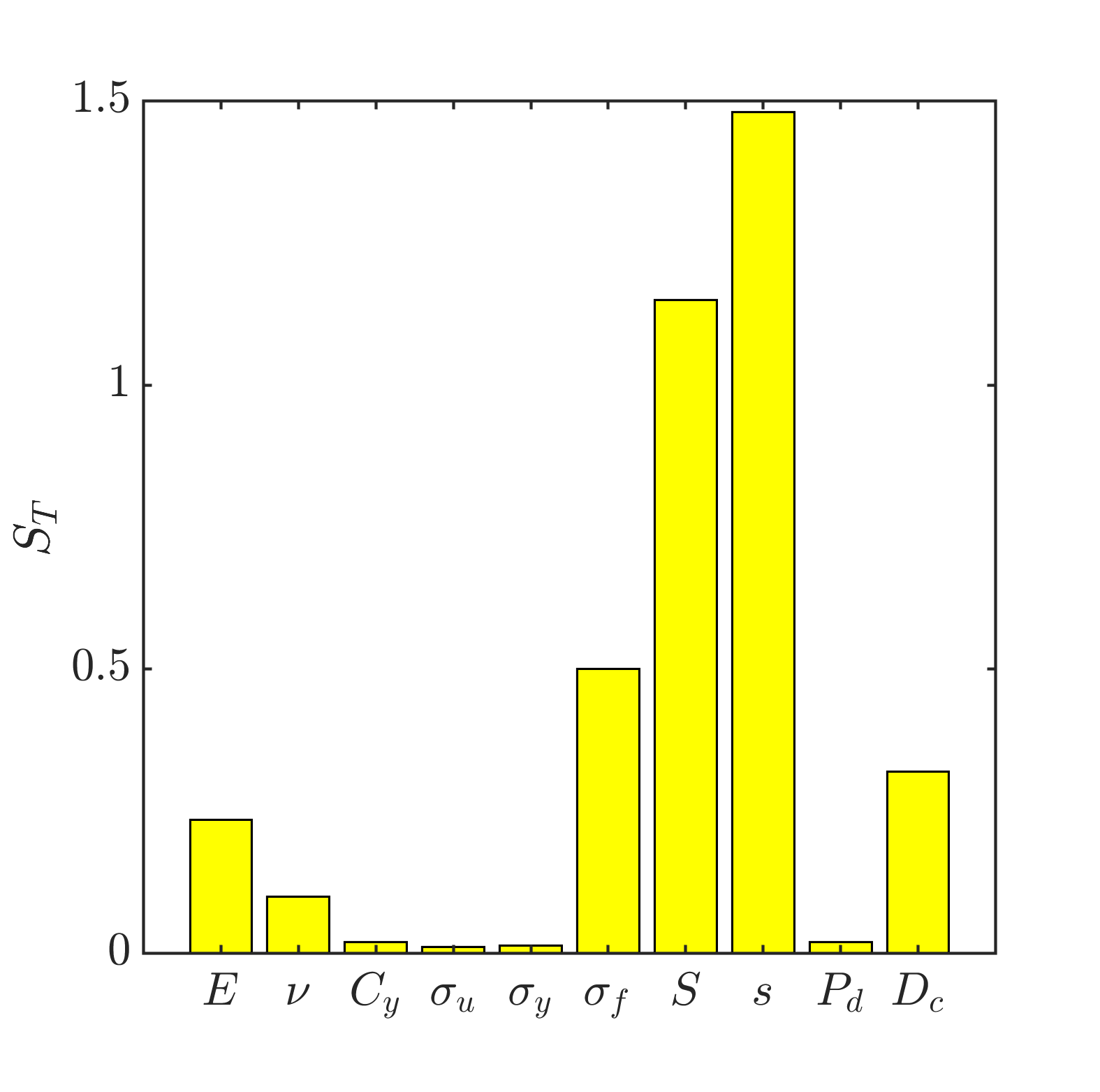}
\end{center}
\vspace{-10mm}
\caption{
Total effect sensitivity index computed using 1000 Monte Carlo samples.
}
\label{fig:total_index}
\end{figure}

The above mentioned method is implemented to estimate total sensitivity indices of the fatigue model parameters and the results are shown in Figure \ref{fig:total_index}. 
\blue{To compute sensitivity indices in this figure 1000 Monte Carlo samples of parameters are used based upon the observed rate of convergence of the indices with respect to number of samples.
Helton and Davis \cite{helton2003latin} have shown using a Latin Hypercube Sample \cite{mckay1979comparison} increases the accuracy of the sensitivity indices.}

It can be seen qualitatively from scatterplots of Figures \ref{fig:scatter1} and \ref{fig:scatter2} and quantitatively from estimated total sensitivity indices of Figure \ref{fig:total_index} that the parameters $S$ and $s$ are the most important
contributors to the fatigue life due to higher values of $\mathcal{S}_T$ and exhibiting distinct patterns of the model output clouds in the scatterplots.
However, other parameters do not show a significant effect on variation of fatigue life of the material.

To verify the results of global sensitivity analyses, the fatigue life is computed considering three sets of parameters,
\textit{first} in which all the parameters are assumed to be random variables sampled from the distributions in Table \ref{table:param_sen},
\textit{second} in which most influential parameters $S$ and $s$ are assumed to be deterministic, and
\textit{third} in which less effective parameters $E$ and $\nu$ are considered as deterministic values.
In the two latter cases the deterministic parameters are excluded from sampling and are fixed to the mean values of the corresponding distributions in Table \ref{table:param_sen}.

\begin{figure}[h]
\begin{center}
\includegraphics[trim = 0mm 0mm 0mm 0mm, clip,width=0.6\textwidth]{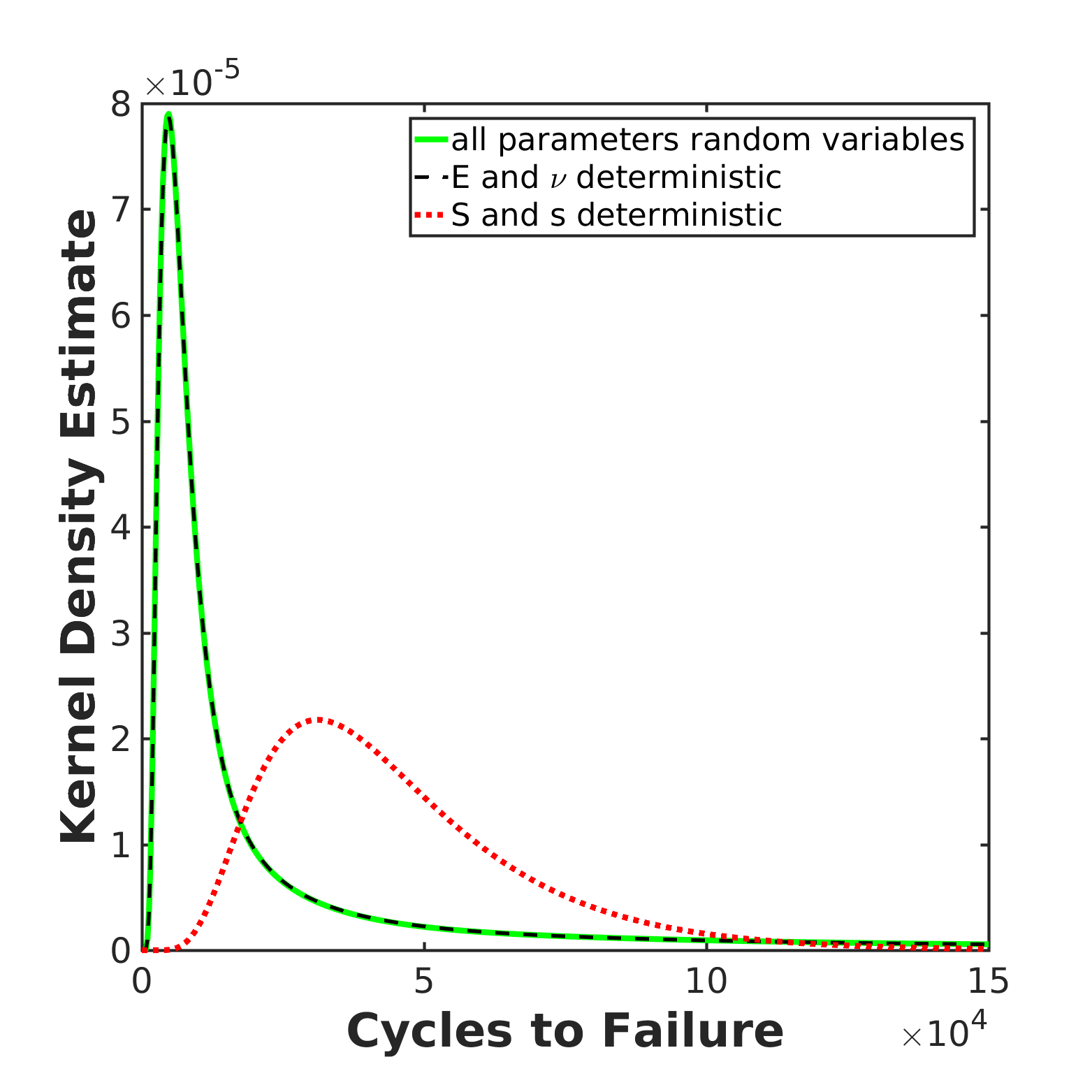}
\end{center}
\vspace{-10mm}
\caption{
Comparison between the kernel density estimates of the number of cycles before failure
resulting from fatigue damage model for the cases when $(S,s)$ and $(E,\nu)$ considered as single (deterministic) values. Results obtained from 2000 Monte Carlo samples of the parameter space according to the distributions presented in Table \ref{table:param_sen} .
}
\label{fig:kde_compare}
\end{figure}

The kernel density estimate (KDE) for each case is presented in Figure \ref{fig:kde_compare}. The minimal difference between the KDEs due to including or excluding $(E,\nu)$ as random variables indicates the insignificant effect of these parameters on fatigue life of the material. However, considering $(S,s)$ as deterministic values results in major changes in KDE confirming the importance of these parameters in fatigue life prediction and consequently hardware design.

Considering the computational cost of solving statistical inverse problem as well as curse of dimensionality, global sensitivity analysis can aid in determining the parameters that can be considered as deterministic during statistical model calibration and greatly reduces computational effort in calibration stage. Given the sensitivity results obtained from fatigue model, we consider only $S$ and $s$ as stochastic parameters in the course of statistical calibration of damage model against fatigue test measurements while other parameters are assumed to be deterministic values. Obviously, deciding on which parameters to be excluded from the calibration process is a subjective decision made by designer for a particular problem.

%

\subsection{Experimental Observations from Fatigue Tests}

\begin{figure}[h!]
\begin{center}
\includegraphics[trim = 7mm 28mm 5mm 10mm, clip,width=0.55\textwidth]{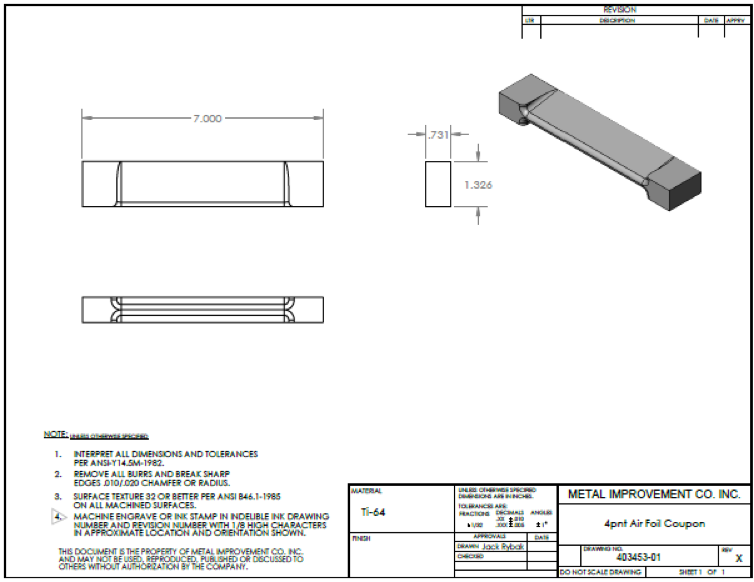}
~
\includegraphics[trim = 50mm 45mm 50mm 125mm, clip,width=0.4\textwidth]{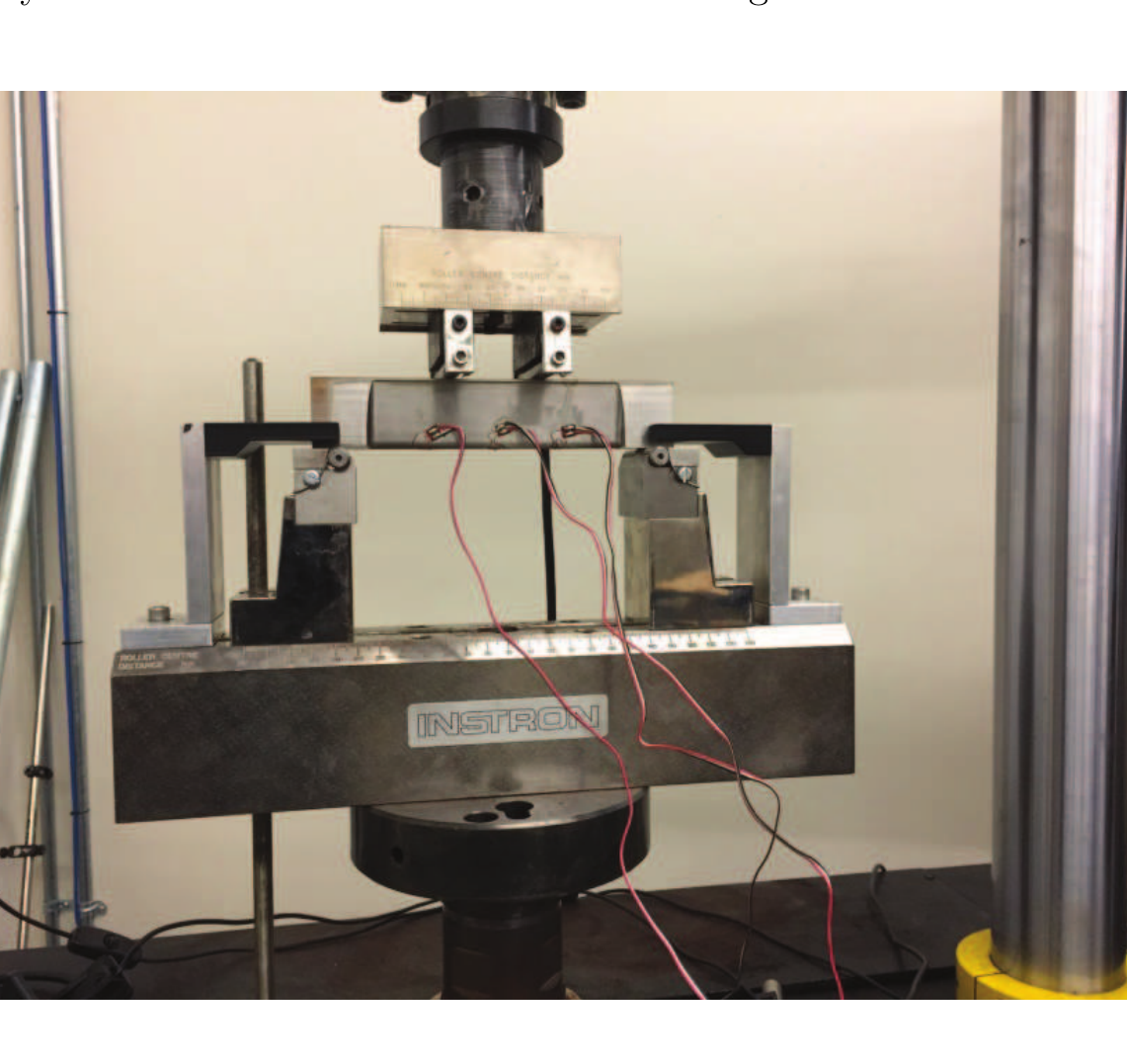}
\\(a) \hspace{70mm} (b)
\end{center}
\vspace{-6mm}
\caption{
(a) Airfoil coupon of Ti6Al4V designed to generate fatigue test data used to calibrate lifetime predictions.
(b) Fatigue test coupon instrumented for strain measurements on fatigue rig.
}
\label{fig:fatigue_test}
\end{figure}

In order to provide data for calibrating the fatigue damage model under uncertainty, fatigue tests were performed on machined airfoil coupons.  In this regard, a 4-point bend coupons of 7 inch length and 1.326 inch height and 0.75 inch base thickness with airfoil contour was designed with geometry as shown in Figure \ref{fig:fatigue_test}(a). Ten coupons were fabricated of the same Ti6Al4V plate material 
(AMS 4928V, flat bar stock) post annealed for 2 hours at 1300$^{\circ}$F.
Curtiss-Wright has built and tested similar 4-point bend coupons for jet engine development applications and had confidence in the reliability of this design.  
Curtiss-Wright Metal Improvement Company (MIC) carried out the fatigue tests.

For fatigue testing, the coupons were loaded into a 4-point bend fixture on an Instron 20 kip test rig as shown in Figure  \ref{fig:fatigue_test}(b).  Stress loading were provided from the actuator and lower load points using 6 inch roller spacing and reacted above by the two rollers with 1.75 inch spacing.  Consequently the region of the bottom edge of the coupon along the center 1.75 inch span received a uniform stress loading during test.  Because this test concept creates a very uniform stress along this edge it allows accurate calibration of the stress loading.

An un-notched coupon was affixed with strain gauges as shown in Figure \ref{fig:fatigue_test}(b) and installed on the Instron 20-kip  fatigue test rig.  Known loads were slowly applied and the strain developed was compared to that predicted from the known modulus of elasticity for the material and the finite element analysis of the coupon design.  The strain gauge data matched at about the 98.5\% level with the finite element prediction, inline with what has been typically seen previously for this kind of test.

\begin{figure}[h!]
\begin{center}
\includegraphics[trim = 0mm 1mm 0mm 0mm, clip,width=0.65\textwidth]{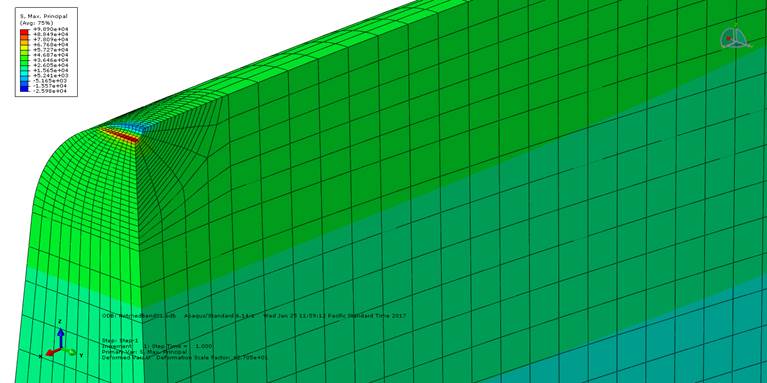}
~
\includegraphics[trim = 0mm 1mm 0mm 0mm, clip,width=0.3\textwidth]{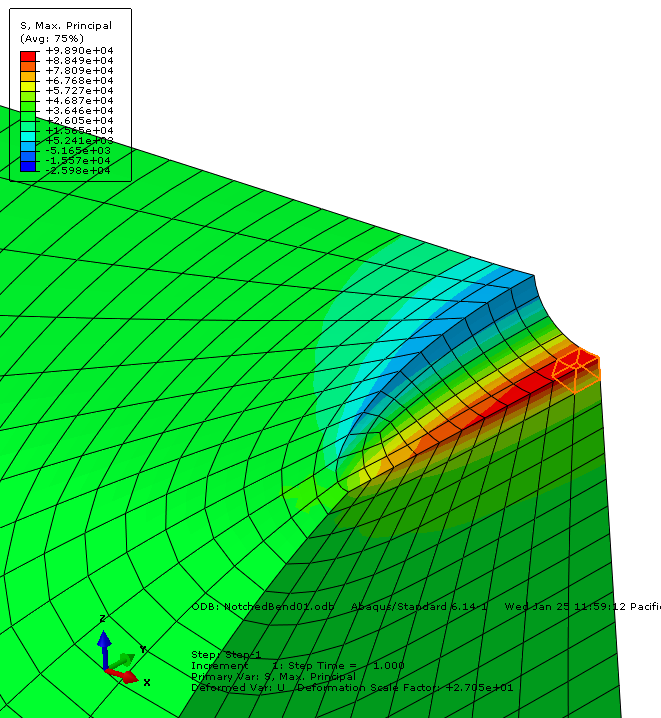}
\end{center}
\vspace{-6mm}
\caption{
Stress analysis of fatigue test coupon including notch agrees with the 35 ksi loading and predicts 99 ksi load in the notch..
}
\label{fig:fem}
\end{figure}

Because this research effort is focused on computational prediction of cyclic fatigue lifetime, which includes crack initiation and crack growth rates, it was decided to initiate a deterministic starting flaw in the coupon leading edge so that the damage initiation and propagation is confined to a local region and it can be tracked easily.
To this end, a notch representing a $K_t$ stress riser was generated in the curved edge of the test coupons using electro discharge machining (EDM).  The notch was cut to 0.010 inches wide by 0.010 inches deep.
With a load of 3645 pounds on the 4-point bend setup, the smooth stress outside of the notch was predicted to be 35 ksi.  The elastically predicted stress for a 0.010" radius notch is 99 ksi, as shown in the analysis line-out of Figure \ref{fig:fem} with the notch at the upper left corner of the model with two planes of symmetry.

Curtiss-Wright used their archived fatigue test data from similar Ti6Al4V coupons to propose an initial test run at 35 ksi stress loading to achieve a targeted run of 200000 cycles.
\blue{Table \ref{tabel:test_data} summarizes the experimental data conducted on five specimens and  will be used for fatigue model calibration.}

\begin{table}[h!]
\centering
\begin{tabular}{lccc}
\hline
Sample \#     & Cycles            & Stress outside of notch & Maximum stress at notch \\ \hline
TDA-6         & 114316            & 45 ksi                    & 111.76 ksi                \\
TDA-2         & 98774             & 45 ksi                    & 111.76 ksi                \\
TDA-4         & 279931            & 35 ksi                    & 87.3154 ksi                \\
TDA-8         & 10441             & 97.8 ksi                  & 243.61 ksi                \\
TDA-10        & 10808             & 97.8 ksi                  & 243.61 ksi                \\ \hline
\end{tabular}
\caption{Tabular data for fatigue test results of Ti6Al4V unpeened coupons.}
\label{tabel:test_data}
\end{table}

\subsection{Bayesian calibration of fatigue model}

Conducting the statistical calibration of damage model against fatigue test experimental observations requires modeling decision on
the form of likelihood function, 
model parameters that are to be treated as random variables (instead of deterministic), and 
prior PDFs that are to be used for the random parameters.
Following the results of global sensitivity presented in the proceeding section, the statistical calibration is performed on two most influential model parameters on fatigue life prediction that controls damage growth, $S$ and $s$, while others are fixed to the mean values reported in the literature for Ti6Al4V material.

Although the fatigue damage forward model considered here can be computed in
a few minutes when set within a statistical inversion computational framework (where the forward problem may
be run hundreds of thousands of times), the required computational effort is not trivial.
Thus, in order to keep the focus on the feasibility of the probabilistic design approach, we limit ourselves to calibrating only two parameters. In this regard, uniform prior PDFs are taken into account for the random parameters $\boldsymbol{\theta} = (S,s)$ over appropriate range (Table \ref{table:param_sen}).
The form of the likelihood function reflects the way the discrepancy between
the quantities computed with the material constitutive relation and the reference data are modeled.
As indicated previously, the reference data $\mathbf{D}$ provided through the fatigue test consist of the measured number of cycles before specimen failure under cyclic loading. It is known that such data is contaminated with noise and specimen variability (i.e., noise in data).
Moreover, fatigue test results are only available in three stress conditions since experimentally investigation of the full spectrum of stress-life is not feasible (i.e., incomplete data).
On the other hand, the continuum damage model is imperfect and is a simplified characterization of reality. For example, the current fatigue computational model does not address microstructural evolution such as density of voids and inclusions (i.e., error in model).

One can assume that the error in observational data and model prediction $\mathcal{E}$ is additive and represented by a Gaussian random variable with zero mean and variance $\Sigma$,
\begin{equation}
\mathcal{E} \sim \mathcal{N}(\mathbf{0}, \Sigma^2\mathbf{I}).
\end{equation}
Following the above consideration along with (\ref{eq:likelihood}), the form of likelihood can be postulated as
\begin{equation}
\ln(\pi_{\rm like}(\mathbf{D}|\boldsymbol{\theta})) =
\frac{n}{2}\ln(2\pi) -  \sum_{k=1}^n \left[\ln(\Sigma_k) +
\frac{D_k - Y(\boldsymbol{\theta}_k)}{2\Sigma_k}
\right]^2,
\end{equation}
where $n$ is number of experimental data.

The statistical inverse problem for Bayesian calibration of fatigue damage model is conducted using the parallel MPI/C++ software libraries (Quantification of Uncertainty for Estimation, Simulation and Optimization - QUESO) \cite{queso}.
QUESO make use of Markov Chain Monte Carlo (MCMC) algorithms for sampling posterior PDFs, following Bayes' formula. 
\blue{
In particular, multilevel sampling (\textit{parallel tempering}) algorithm \cite{prudencio2012multi} is employed for statistical calibration of fatigue model.
This algorithm improves the chances of exploring all existing modes of the posterior PDF through sampling increasingly difficult
intermediate distributions, accumulating information from one intermediate
distribution to the next, until the target posterior distribution is better sampled. Possible intermediate distributions are given by 
\begin{equation}\label{eq-intermidiate}
\pi^{(\ell)}_{\text{int}}({\boldsymbol{\theta}}|\mathbf{D})
=
\left[
\pi_{\text{like}}(\mathbf{D}|{\boldsymbol{\theta}})
\right]^{\alpha_{\ell}}
.
\pi_{\text{prior}}({\boldsymbol{\theta}}),
\qquad
\ell = 0, 1, \cdots, L,
\end{equation}
for a given $ L>0$ and a sequence $0=\alpha_0 < \alpha_1 < \ldots < \alpha_L=1 $ where $\alpha_\ell=0$ and $\alpha_\ell=1$ denote the prior and posterior distribution respectively. Therefore as $\ell$ increases, the distribution transitions
from the initial prior to the (eventually multimodal) posterior \cite{prudencio2012multi}.
Such sampling algorithms can greatly benefit from parallel computing.
}

\begin{figure}[h]
\begin{center}
\includegraphics[trim = 0mm 0mm 10mm 0mm, clip,width=0.34\textheight]{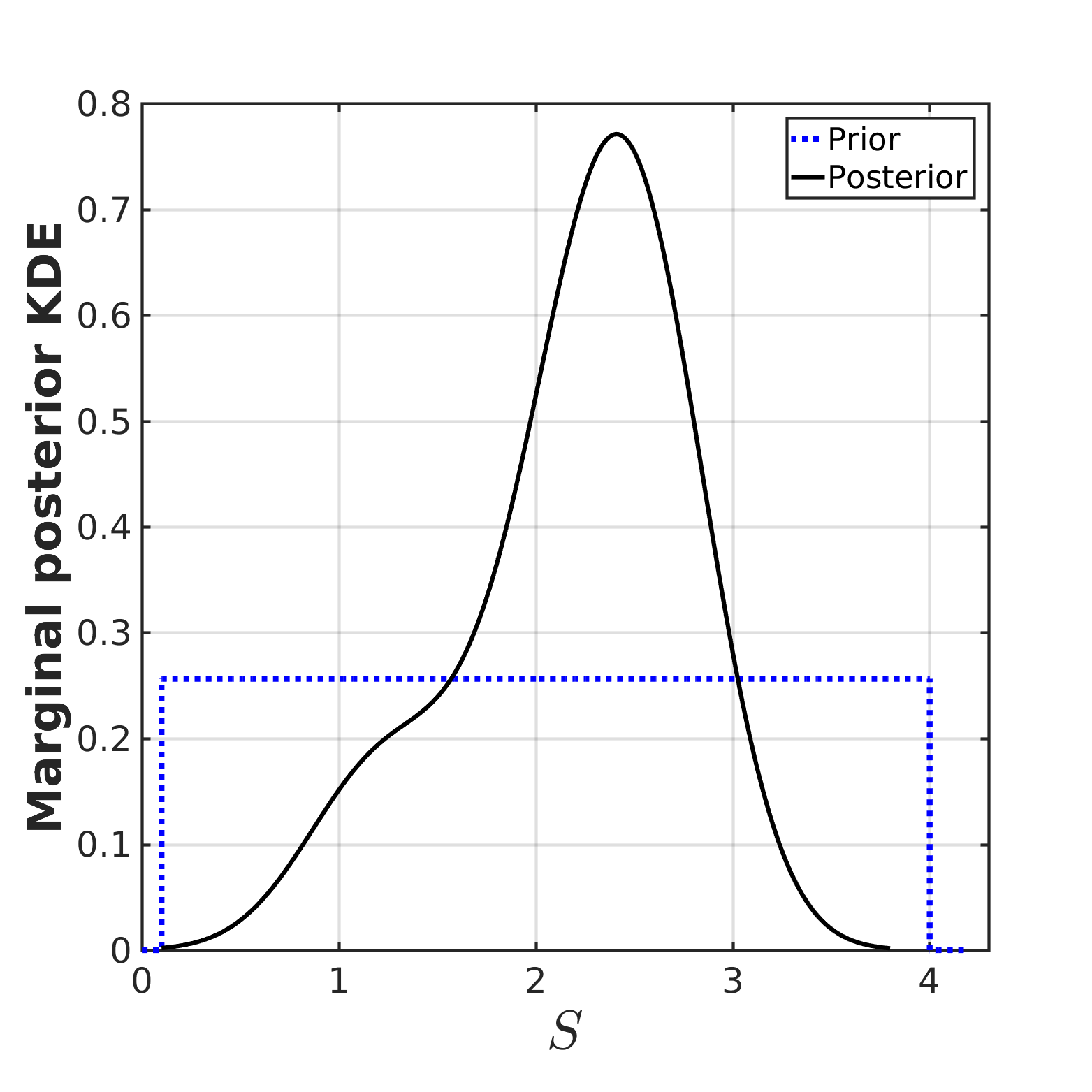}
~
\includegraphics[trim = 0mm 0mm 10mm 0mm, clip,width=0.34\textheight]{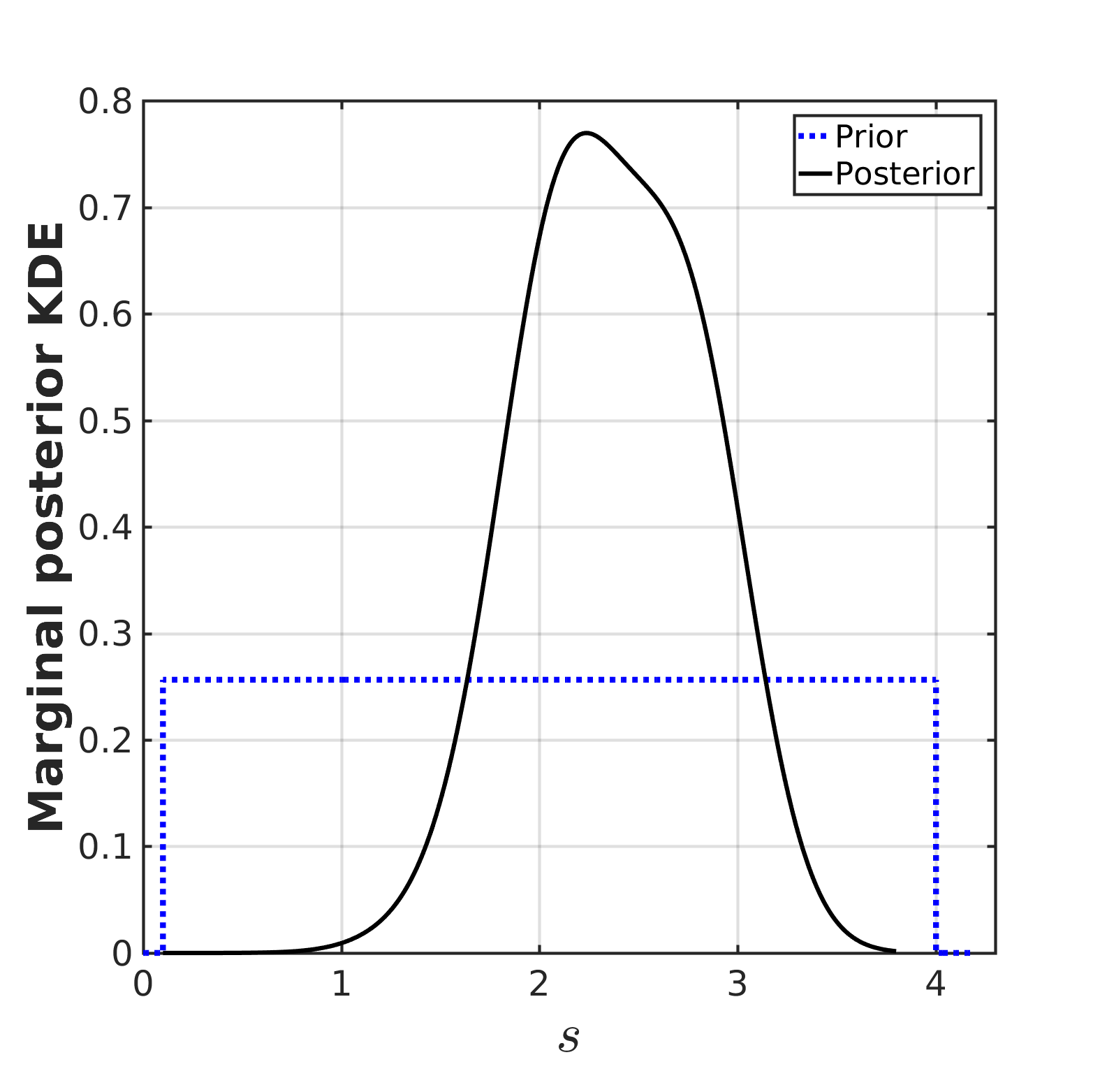}
\\(a) \hspace{70mm} (b)
\end{center}
\vspace{-6mm}
\caption{
Calibrated material parameters of the fatigue damage model. Posterior marginal density estimation
of (a) damage denominator $S$; (b) damage exponent $s$.
}
\label{fig:pdfs_Ss}
\end{figure}

Figure \ref{fig:pdfs_Ss} shows the computed posterior marginal kernel density estimation (KDE) of the parameters $(S,s)$ as a result of calibrating fatigue model against given 5 experimental measurements (Table \ref{tabel:test_data}). Wide distributions obtained for the inferred parameters indicate large amount of uncertainty in both experiment (noise and incompleteness of data) and the way we modeled the fatigue damage (constitutive model inadequacy, modeling assumption, assessing stress at the notch). Deterministic parameter calibration is also conducted based on least squares regression and  direct search method for multidimensional unconstrained minimization based on Nelder-Mead simplex algorithm \cite{lagarias1998}. The deterministic calibration of fatigue model resulted in $S$ = 2.2413 and $s$ = 2.2451. Despite inexpensive computational cost of evaluating these deterministic values, such parameters might result in incorrect fatigue life prediction since they do not carry information regarding the uncertainty in model prediction.

%

\subsection{Probabilistic design of fatigue life}

Final stage in predictive modeling involves employment of the calibrated model with quantified uncertainties to make predictions as well as design decisions.
This will be addressed in this section for fatigue life prediction of metallic component using the statistically calibrated fatigue damage model.

\begin{figure}[h]
\begin{center}
\includegraphics[trim = 0mm 0mm 0mm 0mm, clip,width=0.45\textheight]{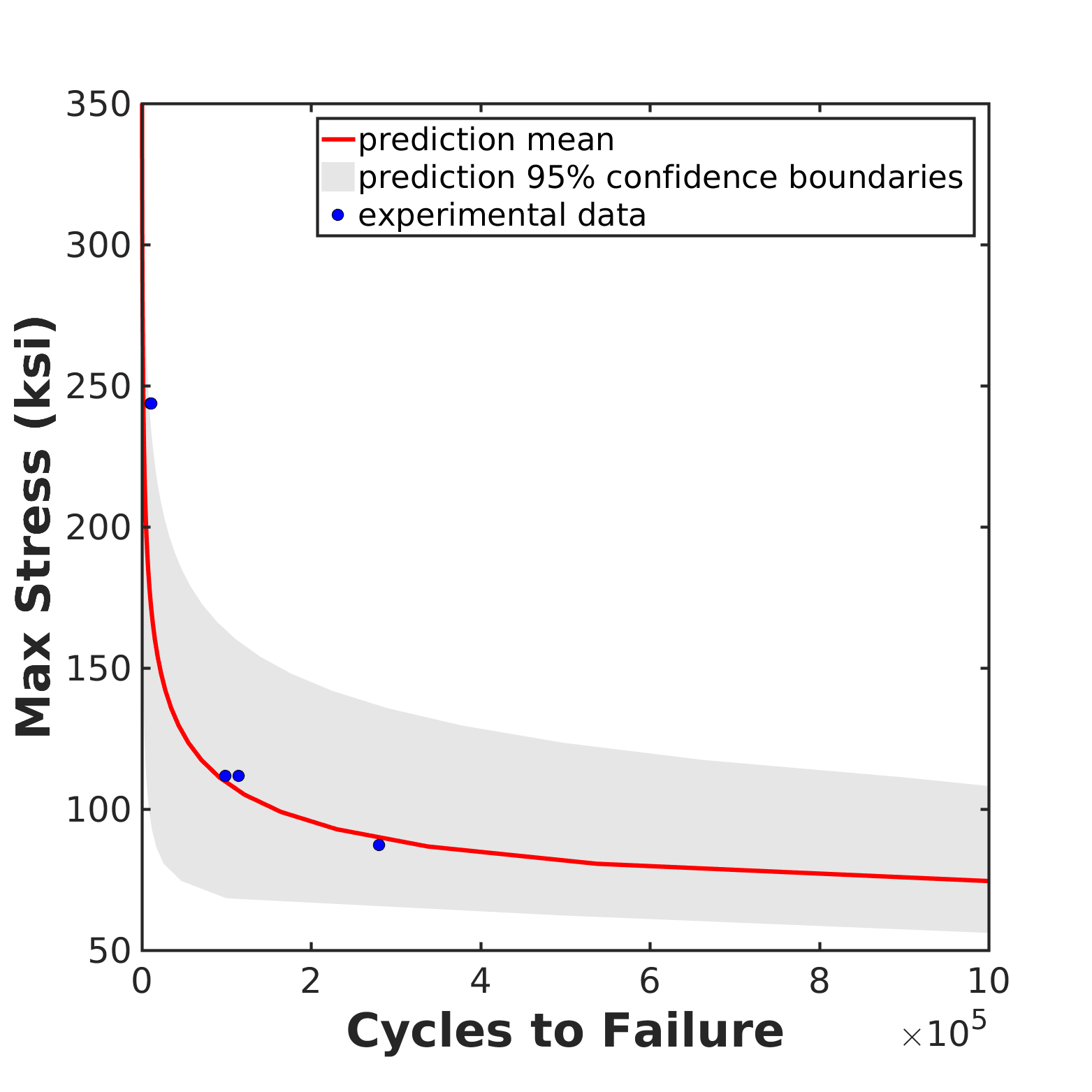}
\end{center}
\vspace{-6mm}
\caption{
Model prediction of fatigue strength of material (S-N diagram) with R = 0.1: mean values and  95\% confidence interval.
}
\label{fig:SNcurve}
\end{figure}

The predicted fatigue strength of material using computational model for wide range of applied load is shown in Figure \ref{fig:SNcurve}.
This figure shows S-N diagram that plots maximum applied stress amplitude versus cycles to
failure for stress ratio of R = 0.1. The mean stress-life values as well as 95\% confidence interval is shown in this figure. The confidence intervals are defined as smallest interval at each stress level where failure cycles have probability of 0.95.
\blue{
The experimental data of the fatigue tests utilized for Bayesian model calibration are also shown in this figure.
The mean values of model prediction show larger discrepancy with the data points at higher stress, while 95\% confidence bound has larger variation in fatigue lives at low stress regimes.
This figure shows although the mean prediction curve (corresponding to deterministic values of calibrated parameters) is close to the experimental observations, there is a significant uncertainty in predicted fatigue lives, i.e. wide confidence intervals in a particular stress level.
This indicates deterministic model calibration is inadequate to deal with the significant uncertainties in model prediction and results in design with unknown reliability.
}

\begin{figure}[h]
\begin{center}
\includegraphics[trim = 0mm 0mm 10mm 0mm, clip,width=0.34\textheight]{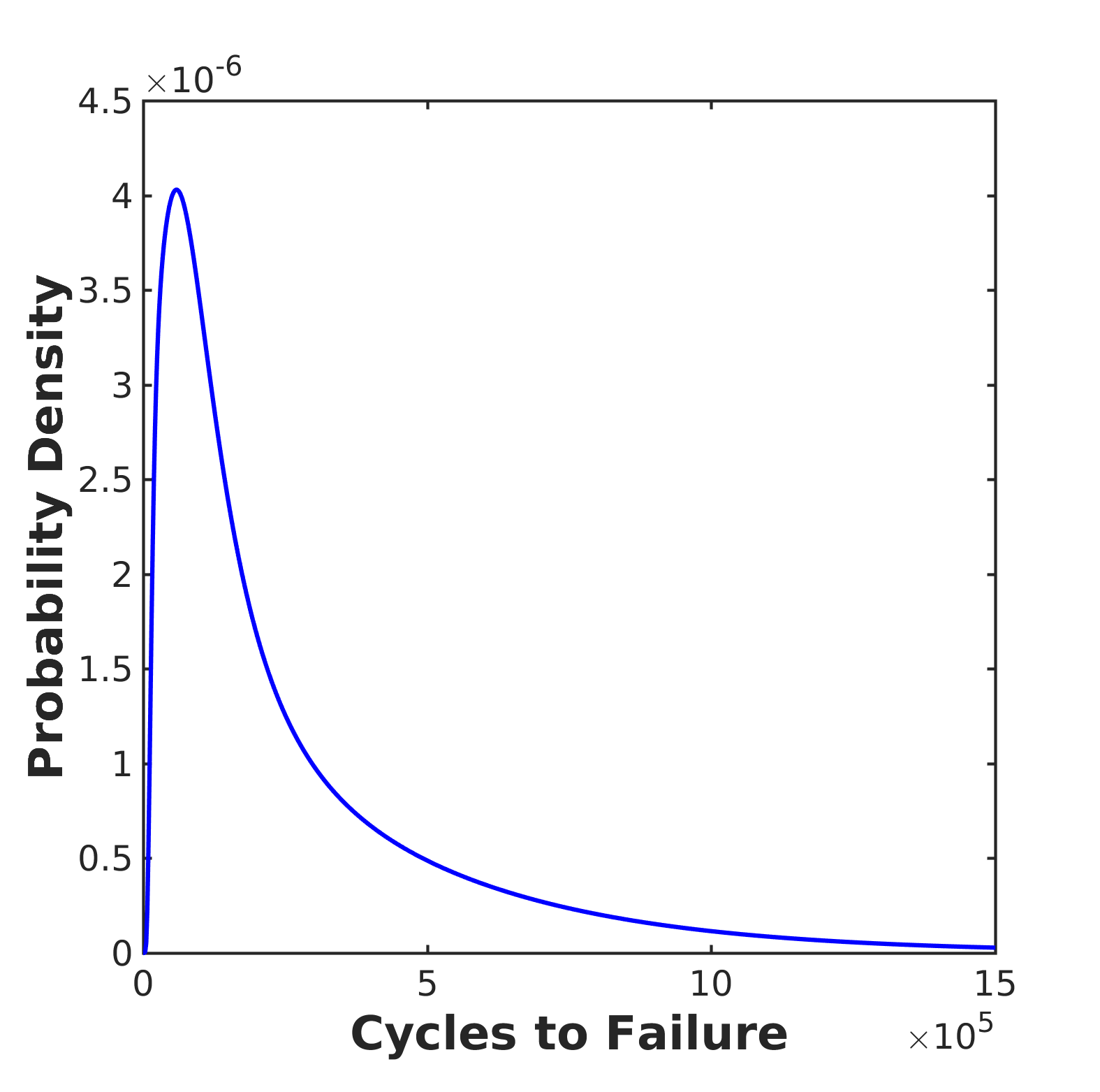}
~
\includegraphics[trim = 0mm 0mm 10mm 0mm, clip,width=0.34\textheight]{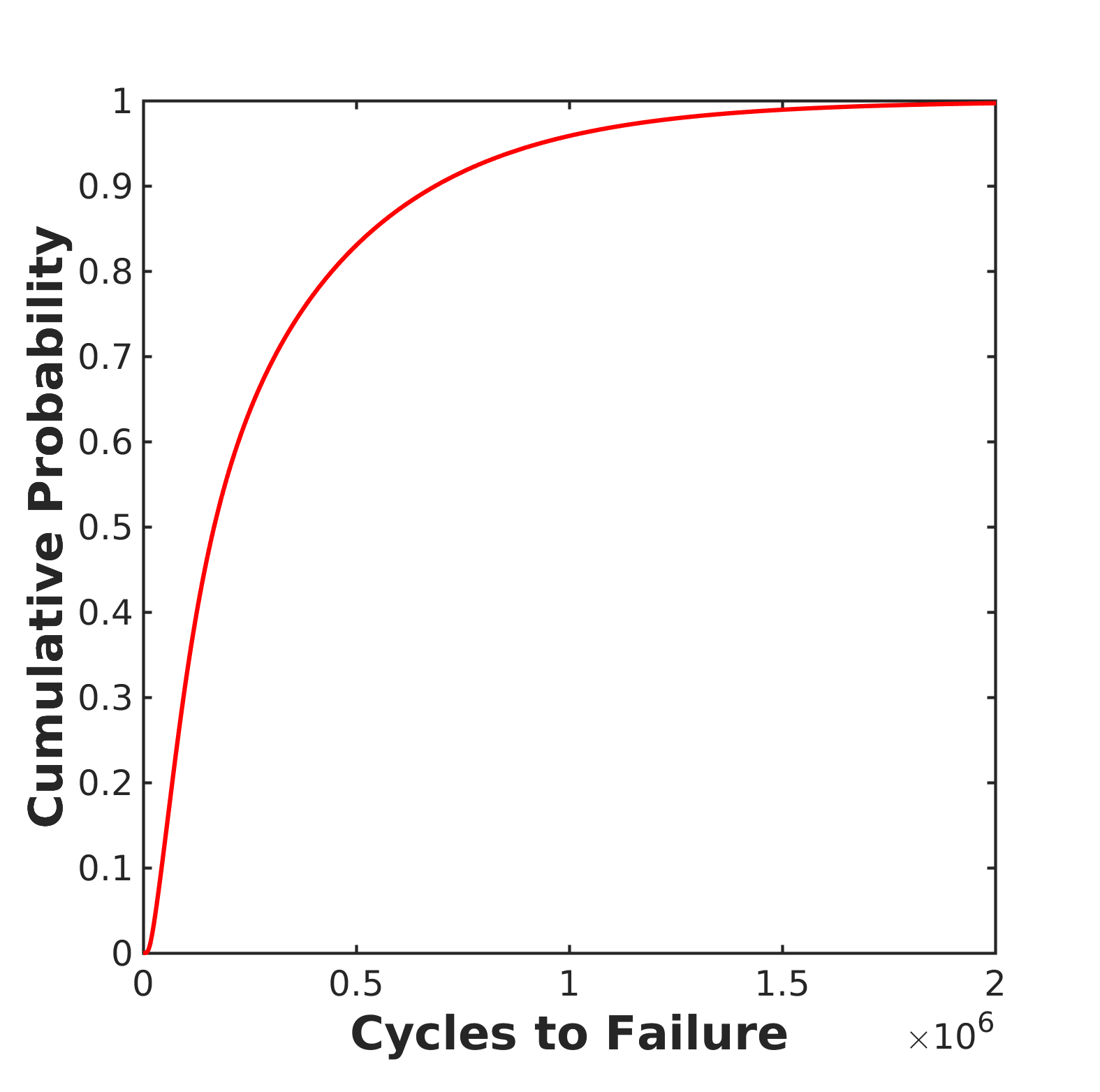}
\\(a) \hspace{70mm} (b)
\end{center}
\vspace{-6mm}
\caption{
Probability density of the failure cycles under maximum stress of 100ksi and R = 0.1 applied load and compressive residual plastic strain of 0.193. Results obtained from 100 Monte Carlo samples of posterior distributions of random parameters.
}
\label{fig:Nf_pdf_cdf}
\end{figure}

As mentioned previously, surface treatment such as laser peening enables increasing fatigue lifetime of a metallic component. The effect of such treatment is included into the fatigue model through compressive residual plastic strain. The amount of residual strain can be controlled through laser peening and thus can be considered as the design variable. In order to investigate the effect of uncertainty in predicting life of a component, a design scenario is considered here as:
\textit{For a given cyclic load, predict optimal residual strain (indicating surface treatment) such that, for a specific probability, the fatigue life of the component falls beyond a specified target number of cycles.}
For illustration it is assumed that the desired (target) performance of the metallic hardware is 50000 cycles. The probability density $\pi(N_f|\boldsymbol{\epsilon}^{rp})$ and the cumulative distribution $\Pi(N_f|\boldsymbol{\epsilon}^{rp})$ are shown in Figure \ref{fig:Nf_pdf_cdf} for maximum stress of 100ksi and R=0.1 condition. Simple iterative method results in minimum effective residual plastic strain of $\epsilon^{rp}_{0}$ = 0.193 to satisfy $\Pi(N_f|\boldsymbol{\epsilon}^{rp})$ = 0.15. In other word, in this case with 85\% of reliability the fatigue life is 50000 cycles.
Note that the applied cyclic stress in this design scenario is different from the stress level of the available fatigue tests. As mentioned previously, in most cases, computational prediction is conducted
 out of the range of available observations.

%

\section{Conclusions}\label{sec:conclusions}

\blue{
In the present study, a new probabilistic design methodology is demonstrated that allows the quantification of uncertainties in experimental data and computational model for fatigue life prediction of structural components as well as propagation of these uncertainties on prediction of components behavior. 
Such methodology integrates the mathematical model of fatigue damage material behavior and the experimental observations with a Bayesian framework for model calibration and uncertainty quantification. 
This enables assessing the level of confidence in computational model prediction to improve design and performance of hardware components beyond the range of test data.
The physical problem under study is the behavior of airfoil coupons of Ti6Al4V material that are common in turbine engines of aircraft structures under cyclic loading. The experimental data in term of number of cycles to failure at different stress levels is obtained from fatigue tests conducted on machined airfoil coupons. A fatigue model based on continuum damage theory is used to model the onset and evolution of damage as well as to forecast the lifetime of the material experiencing cyclic loading.
}

\blue{
A computationally feasible framework is developed and implemented in this study to quantify the uncertainty in experimental data due to incomplete and noisy measurements, modeling error due to constitutive model inadequacy, and other modeling assumptions typically made in fatigue life prediction. 
The framework depicted in Figure \ref{fig:flowchart} involves global sensitivity analysis to determine the important model parameters, Bayesian calibration of model against fatigue test data with quantified uncertainty in model and data, and predicting fatigue strength of material as well as design decision making under uncertainty. An illustrative inverse design problem is presented to compute the optimal residual plastic strain, which is a controllable variable through surface treatment, based on desired component performance and expected reliability.
}

\blue{
The outcome of integration among experimental data, computational fatigue damage model,
and statistical analyses indicates presence of large uncertainty in the prediction delivered by the model. This proves the necessity of conducting probabilistic analysis and design for cyclic fatigue life of metallic components.
Presence of such uncertainty is mainly due to modeling assumptions as well as lack of experimental data at wide range of stress levels. It is expected using micromechanical based model that account for evolution of micro-crack and voids, employing finite element model of the full coupon instead of analyzing a hot spot at the notch, and obtaining more experimental data at different stress level will reduce the uncertainties and increase level of confidence regarding the hardware performance and design in practical applications.
}
\blue{
Moreover, the general probabilistic design framework proposed in this work enables accounting for other sources of uncertainties. For example, fatigue tests data on peened specimen provides opportunity for statistical calibration of residual plastic strain input variable of the model. Accounting for the inevitable uncertainty due to surface treatments and propagation of such variability to the life prediction of a component, greatly increases the benefit of these methods to extend component lifetimes as well as reliable assessment in structural safety enhancement.
}
%

\section*{Acknowledgments}
The work reported here was supported by NAVAIR under SBIR contract No.
N68335-16-C-0516.

\bibliographystyle{abbrv}  
\bibliography{refs}

\begin{thebibliography}{10}

\bibitem{allahverdizadeh2012}
N.~Allahverdizadeh, A.~Manes, and M.~Giglio.
\newblock Identification of damage parameters for ti-6al-4v titanium alloy
  using continuum damage mechanics.
\newblock {\em Materialwissenschaft und Werkstofftechnik}, 43(5):435--440,
  2012.

\bibitem{babuvska2016fatigue}
I.~Babu{\v{s}}ka, Z.~Sawlan, M.~Scavino, B.~Szab{\'o}, and R.~Tempone.
\newblock Bayesian inference and model comparison for metallic fatigue data.
\newblock {\em Computer Methods in Applied Mechanics and Engineering},
  304:171--196, 2016.

\bibitem{bahloul2016probabilistic}
A.~Bahloul, A.~B. Ahmed, M.~Mhala, and C.~Bouraoui.
\newblock Probabilistic approach for predicting fatigue life improvement of
  cracked structure repaired by high interference fit bushing.
\newblock {\em The International Journal of Advanced Manufacturing Technology},
  pages 1--13, 2016.

\bibitem{birnbaum1958}
Z.~Birnbaum and S.~C. Saunders.
\newblock A statistical model for life-length of materials.
\newblock {\em Journal of the American Statistical Association},
  53(281):151--160, 1958.

\bibitem{bolotin2001}
V.~Bolotin and I.~Belousov.
\newblock Early fatigue crack growth as the damage accumulation process.
\newblock {\em Probabilistic engineering mechanics}, 16(4):279--287, 2001.

\bibitem{somersalo1}
D.~Calvetti and E.~Somersalo.
\newblock {\em Introduction to Bayesian Scientific Computing: Ten Lectures on
  Subjective Computing}.
\newblock Springer, 2007.

\bibitem{correia2017}
J.~Correia, N.~Apetre, A.~Arcari, A.~De~Jesus, M.~Mu{\~n}iz-Calvente,
  R.~Cal{\c{c}}ada, F.~Berto, and A.~Fern{\'a}ndez-Canteli.
\newblock Generalized probabilistic model allowing for various fatigue damage
  variables.
\newblock {\em International Journal of Fatigue}, 100:187--194, 2017.

\bibitem{Cukier1973}
R.~I. Cukier, C.~M. Fortuin, K.~E. Shuler, A.~G. Petschek, and J.~H. Schaibly.
\newblock Study of the sensitivity of coupled reaction systems to uncertainties
  in rate coefficients. i theory.
\newblock {\em The Journal of Chemical Physics}, 59(8):3873--3878, 1973.

\bibitem{dingreville2010}
R.~Dingreville, C.~C. Battaile, L.~N. Brewer, E.~A. Holm, and B.~L. Boyce.
\newblock The effect of microstructural representation on simulations of
  microplastic ratcheting.
\newblock {\em International Journal of Plasticity}, 26(5):617--633, 2010.

\bibitem{Eshelby}
J.~D. Eshelby.
\newblock The determination of the elastic field of an ellipsoidal inclusion,
  related problems.
\newblock {\em Proceedings of the Royal Society of London. Series A,
  Mathematical and Physical Sciences}, 241(1226):376--396, 1957.

\bibitem{farrell2014}
K.~Farrell and J.~T. Oden.
\newblock Calibration and validation of coarse-grained models of atomic
  systems: Application to semiconductor manufacturing.
\newblock {\em Comput. Mech.}, 54(1):3--19, July 2014.

\bibitem{Farrell2015}
K.~Farrell, J.~T. Oden, and D.~Faghihi.
\newblock A bayesian framework for adaptive selection, calibration, and
  validation of coarse-grained models of atomistic systems.
\newblock {\em Journal of Computational Physics}, 295:189 -- 208, 2015.

\bibitem{hammer2012}
J.~T. Hammer.
\newblock {\em Plastic deformation and ductile fracture of Ti-6Al-4V under
  various loading conditions}.
\newblock PhD thesis, The Ohio State University, 2012.

\bibitem{helton2003latin}
J.~C. Helton and F.~J. Davis.
\newblock Latin hypercube sampling and the propagation of uncertainty in
  analyses of complex systems.
\newblock {\em Reliability Engineering \& System Safety}, 81(1):23--69, 2003.

\bibitem{HommaSaltelli1996}
T.~Homma and A.~Saltelli.
\newblock Importance measures in global sensitivity analysis of nonlinear
  models.
\newblock {\em Reliability Engineering and System Safety}, 52(1):1 -- 17, 1996.

\bibitem{kaipio2006}
J.~Kaipio and E.~Somersalo.
\newblock {\em Statistical and computational inverse problems}, volume 160.
\newblock Springer Science \& Business Media, 2006.

\bibitem{Kroner}
E.~Kroner.
\newblock On the plastic deformation of polycrystals.
\newblock {\em Acta Metall.}, 9:155–161, 1961.

\bibitem{kwon2011probabilistic}
K.~Kwon, D.~M. Frangopol, and M.~Soliman.
\newblock Probabilistic fatigue life estimation of steel bridges by using a
  bilinear s-n approach.
\newblock {\em Journal of Bridge Engineering}, 17(1):58--70, 2011.

\bibitem{lagarias1998}
J.~C. Lagarias, J.~A. Reeds, M.~H. Wright, and P.~E. Wright.
\newblock Convergence properties of the nelder--mead simplex method in low
  dimensions.
\newblock {\em SIAM Journal on optimization}, 9(1):112--147, 1998.

\bibitem{lardner1967}
R.~Lardner.
\newblock A theory of random fatigue.
\newblock {\em Journal of the Mechanics and Physics of Solids}, 15(3):205--221,
  1967.

\bibitem{lemaitre2012}
J.~Lemaitre.
\newblock {\em A course on damage mechanics}.
\newblock Springer Science \& Business Media, 2012.

\bibitem{lemaitre2005}
J.~Lemaitre and R.~Desmorat.
\newblock {\em Engineering damage mechanics: ductile, creep, fatigue and
  brittle failures}.
\newblock Springer Science \& Business Media, 2005.

\bibitem{lemaitre1999two}
J.~Lemaitre, J.~Sermage, and R.~Desmorat.
\newblock A two scale damage concept applied to fatigue.
\newblock {\em International Journal of fracture}, 97(1-4):67--81, 1999.

\bibitem{mashayekhi2013}
M.~Mashayekhi, A.~Taghipour, A.~Askari, and M.~Farzin.
\newblock Continuum damage mechanics application in low-cycle thermal fatigue.
\newblock {\em International Journal of Damage Mechanics}, 22(2):285--300,
  2013.

\bibitem{mcdowell2010}
D.~McDowell and F.~Dunne.
\newblock Microstructure-sensitive computational modeling of fatigue crack
  formation.
\newblock {\em International journal of fatigue}, 32(9):1521--1542, 2010.

\bibitem{mckay1979comparison}
M.~D. McKay, R.~J. Beckman, and W.~J. Conover.
\newblock Comparison of three methods for selecting values of input variables
  in the analysis of output from a computer code.
\newblock {\em Technometrics}, 21(2):239--245, 1979.

\bibitem{naderi2013}
M.~Naderi, S.~Hoseini, and M.~Khonsari.
\newblock Probabilistic simulation of fatigue damage and life scatter of
  metallic components.
\newblock {\em International Journal of Plasticity}, 43:101--115, 2013.

\bibitem{odenbauskafaghihi}
J.~T. {Oden}, I.~{Babuska}, and D.~{Faghihi}.
\newblock {\em Encyclopedia of Computational Mechanics}, chapter Predictive
  Computational Science: Computer Predictions in the Presence of Uncertainty.
  Eds. Erwin Stein, Rene de Borst and Thomas J.R. Hughes.
\newblock John Wiley \& Sons, Ltd.

\bibitem{OdenMoserGhattas2010I}
T.~Oden, R.~Moser, and O.~Ghattas.
\newblock {Computer Predictions with Quantified Uncertainty, Part I}.
\newblock {\em SIAM News}, 43(9), 2010.

\bibitem{OdenMoserGhattas2010II}
T.~Oden, R.~Moser, and O.~Ghattas.
\newblock {Computer Predictions with Quantified Uncertainty, Part II}.
\newblock {\em SIAM News}, 43(10), 2010.

\bibitem{ortiz1988}
K.~Ortiz and A.~S. Kiremidjian.
\newblock Stochastic modeling of fatigue crack growth.
\newblock {\em Engineering Fracture Mechanics}, 29(3):317--334, 1988.

\bibitem{prudencio2014compB}
E.~Prudencio, P.~Bauman, S.~Williams, D.~Faghihi, K.~Ravi-Chandar, and J.~Oden.
\newblock Real-time inference of stochastic damage in composite materials.
\newblock {\em Composites Part B: Engineering}, 67:209--219, 2014.

\bibitem{prudencio2012multi}
E.~Prudencio and S.~H. Cheung.
\newblock Parallel adaptive multilevel sampling algorithms for the bayesian
  analysis of mathematical models.
\newblock {\em International Journal for Uncertainty Quantification}, 2(3),
  2012.

\bibitem{queso}
E.~Prudencio and K.~Schulz.
\newblock The parallel c++ statistical library ‘queso’: Quantification of
  uncertainty for estimation, simulation and optimization.
\newblock In M.~Alexander, P.~D’Ambra, A.~Belloum, G.~Bosilca, M.~Cannataro,
  M.~Danelutto, B.~Martino, M.~Gerndt, E.~Jeannot, R.~Namyst, J.~Roman,
  S.~Scott, J.~Traff, G.~Vallée, and J.~Weidendorfer, editors, {\em Euro-Par
  2011: Parallel Processing Workshops}, volume 7155 of {\em Lecture Notes in
  Computer Science}, pages 398--407. Springer Berlin Heidelberg, 2012.

\bibitem{prudencio2014ijnme}
E.~E. Prudencio, P.~T. Bauman, D.~Faghihi, K.~Ravi-Chandar, and J.~T. Oden.
\newblock A computational framework for dynamic data-driven material damage
  control, based on bayesian inference and model selection.
\newblock {\em International Journal for Numerical Methods in Engineering},
  2014.

\bibitem{prudencio2013}
E.~E. Prudencio, P.~T. Bauman, S.~Williams, D.~Faghihi, K.~Ravi-Chandar, and
  J.~T. Oden.
\newblock A dynamic data driven application system for real-time monitoring of
  stochastic damage.
\newblock {\em Procedia Computer Science}, 18:2056--2065, 2013.

\bibitem{Saltelli2002}
A.~Saltelli.
\newblock Making best use of model evaluations to compute sensitivity indices.
\newblock {\em Computer Physics Communications}, 145(2):280 -- 297, 2002.

\bibitem{Saltelli2010}
A.~Saltelli, P.~Annoni, I.~Azzini, F.~Campolongo, M.~Ratto, and S.~Tarantola.
\newblock Variance based sensitivity analysis of model output. design and
  estimator for the total sensitivity index.
\newblock {\em Computer Physics Communications}, 181(2):259 -- 270, 2010.

\bibitem{saltelli2008}
A.~Saltelli, M.~Ratto, T.~Andres, F.~Campolongo, J.~Cariboni, D.~Gatelli,
  M.~Saisana, and S.~Tarantola.
\newblock {\em Global Sensitivity Analysis: The Primer}.
\newblock Wiley, 2008.

\bibitem{SaltelliTarantola2002}
A.~Saltelli and S.~Tarantola.
\newblock On the relative importance of input factors in mathematical models.
\newblock {\em Journal of the American Statistical Association},
  97(459):702--709, 2002.

\bibitem{shen2000}
H.~Shen, J.~Lin, and E.~Mu.
\newblock Probabilistic model on stochastic fatigue damage.
\newblock {\em International Journal of Fatigue}, 22(7):569--572, 2000.

\bibitem{Sobol1990}
I.~M. Sobol'.
\newblock Sensitivity estimates for nonlinear mathematical models.
\newblock {\em Matematicheskoe Modelirovanie}, 2:112--118, 1990.

\bibitem{Sobol1993}
I.~M. Sobol'.
\newblock Sensitivity analysis for non-linear mathematical models.
\newblock {\em Mathematical Modeling and Computational Experiment}, 1:407--414,
  1993.

\bibitem{tarantola}
A.~Tarantola.
\newblock {\em Inverse Problem Theory and Methods for Model Parameter
  Estimation}.
\newblock SIAM, 2005.

\bibitem{yang1983}
J.~Yang, G.~Salivar, and C.~Annis.
\newblock Statistical modeling of fatigue-crack growth in a nickel-base
  superalloy.
\newblock {\em Engineering Fracture Mechanics}, 18(2):257--270, 1983.

\bibitem{yatnalkar2010}
R.~S. Yatnalkar.
\newblock {\em Experimental investigation of plastic deformation of ti-6al-4v
  under various loading conditions}.
\newblock PhD thesis, The Ohio State University, 2010.

\bibitem{yu2016}
T.~Yu.
\newblock {\em Continuum damage mechanics models and their applications to
  composite components of aero-engines}.
\newblock PhD thesis, University of Nottingham, 2016.

\bibitem{zherebtsov2005}
S.~Zherebtsov, G.~Salishchev, R.~Galeyev, and K.~Maekawa.
\newblock Mechanical properties of ti--6al--4v titanium alloy with
  submicrocrystalline structure produced by severe plastic deformation.
\newblock {\em Materials Transactions}, 46(9):2020--2025, 2005.

\bibitem{zhu2017probabilistic}
S.~Zhu, S.~Foletti, and S.~Beretta.
\newblock Probabilistic framework for multiaxial lcf assessment under material
  variability.
\newblock {\em International Journal of Fatigue}, 2017.

\bibitem{zhu2015}
S.-P. Zhu, H.-Z. Huang, Y.~Li, Y.~Liu, and Y.~Yang.
\newblock Probabilistic modeling of damage accumulation for time-dependent
  fatigue reliability analysis of railway axle steels.
\newblock {\em Proceedings of the Institution of Mechanical Engineers, Part F:
  Journal of Rail and Rapid Transit}, 229(1):23--33, 2015.

\bibitem{zhu2012}
S.-P. Zhu, H.-Z. Huang, V.~Ontiveros, L.-P. He, and M.~Modarres.
\newblock Probabilistic low cycle fatigue life prediction using an energy-based
  damage parameter and accounting for model uncertainty.
\newblock {\em International Journal of Damage Mechanics}, 21(8):1128--1153,
  2012.

\bibitem{zhu2016}
S.-P. Zhu, H.-Z. Huang, W.~Peng, H.-K. Wang, and S.~Mahadevan.
\newblock Probabilistic physics of failure-based framework for fatigue life
  prediction of aircraft gas turbine discs under uncertainty.
\newblock {\em Reliability Engineering \& System Safety}, 146:1--12, 2016.

\bibitem{zhu2013bayesian}
S.-P. Zhu, H.-Z. Huang, R.~Smith, V.~Ontiveros, L.-P. He, and M.~Modarres.
\newblock Bayesian framework for probabilistic low cycle fatigue life
  prediction and uncertainty modeling of aircraft turbine disk alloys.
\newblock {\em Probabilistic Engineering Mechanics}, 34:114--122, 2013.

\end{thebibliography}
\index{Bibliography@\emph{Bibliography}}%


\end{document}